\documentclass{jfm}
\usepackage{amsfonts,amsmath,amssymb}

\usepackage{multiJFMequations}
\usepackage{accents}
\usepackage{graphicx}
\usepackage[latin1]{inputenc}
\usepackage{mathrsfs}
\usepackage{xcolor}
\usepackage{float}
\usepackage[export]{adjustbox}[2011/08/13]
\usepackage{cancel}
\usepackage{caption}
\usepackage[percent]{overpic}

\def\pd{\partial}

\newcommand*{\be}{\begin{equation}}
\newcommand*{\ee}{\end{equation}}
\newcommand*{\bse}{\begin{subequations}}
\newcommand*{\ese}{\end{subequations}}
\newcommand*{\bme}{\begin{multiequations}}
\newcommand*{\eme}{\end{multiequations}}
\newcommand*{\se}{\singleequation}

\newcommand*{\te}{\tripleequation}

\newcommand {\bal}{\begin{align}}
\newcommand {\eal}{\end{align}}

\renewcommand*{\pd}{\partial}
\newcommand*{\fb}{\frac}

\renewcommand{\Delta}{\triangle}

\newcommand*{\rir}{\right \rangle}
\newcommand*{\lel}{\left \langle}

\renewcommand{\dag}{\dagger}
\newcommand{\vt}{\tilde{\bf u}}
\newcommand{\rt}{\tilde{\rho}}

\newcommand {\tcb}{\textcolor{black}}
\newcommand {\tcr}{\textcolor{black}} 
\makeatletter
\def\mathcolor#1#{\@mathcolor{#1}}
\def\@mathcolor#1#2#3{%
  \protect\leavevmode
  \begingroup
    \color#1{#2}#3%
  \endgroup
}
\makeatother

\begin{document}

\title[Optimal mixing in stratified plane Poiseuille flow]{Optimal mixing in two-dimensional stratified plane Poiseuille flow at finite
P\'eclet  and Richardson numbers}
\author[F. Marcotte \& C.~P.~Caulfield]
{F.~ Marcotte$^{1}$ and C.~P.~Caulfield$^{2,1}$\thanks{Email address for correspondence:
    cpc12@cam.ac.uk}}

\affiliation{$^1$Department of Applied Mathematics and Theoretical Physics, University of Cambridge, Wilberforce Road, Cambridge CB3~0WA, UK\\

$^2$BP Institute, University of Cambridge, Madingley Road, Cambridge CB3 0EZ, UK}

\maketitle

\begin{abstract}
We consider the nonlinear optimisation of irreversible mixing induced
by an initial finite amplitude perturbation of a statically stable
density-stratified fluid with kinematic viscosity $\nu$ and density
diffusivity $\kappa$. The initial
diffusive error function density distribution  varies continuously so that 
\tcb{$\rho \in [\bar{\rho} - \tfrac12\rho_0, \bar{\rho} +  \tfrac12 \rho_0]$}.
A constant pressure gradient is imposed in a plane two-dimensional
channel
of depth $2h$. We 
consider flows with  
a finite P\'eclet number $\Pen= U_m h /\kappa=500$ and Prandtl number
$\Pran=\nu/\kappa=1$,  and a range of bulk Richardson numbers $Ri_b= g
\rho_0 h /(\bar{\rho} U^2) \in [0,1]$
where $U_m$ is the maximum flow speed  of the laminar parallel flow,
and $g$ is the gravitational acceleration.
We  use the constrained variational  direct-adjoint-looping (DAL)
method
to solve two optimization problems, extending the optimal mixing
results of \citet{FCS14} to stratified flows, where
the \tcr{irreversible} mixing of the \tcr{active} scalar density 
\tcr{leads to a conversion of 
kinetic energy into potential energy.}
We identify initial perturbations of fixed finite kinetic energy which
maximize the time-averaged perturbation kinetic energy developed by the
perturbations over a finite time interval, 
 and initial perturbations that minimise the value 
(at a target time, chosen to be $T=10$)
of a   `mix-norm'
 as \tcb{first introduced by \cite{MMP05}, further discussed by \citet{T12}} and shown by 
\cite{FCS14} 
to be
 a computationally efficient and robust proxy for
identifying perturbations
that minimise the 
\tcr{long-time} variance of a scalar
distribution. We demonstrate, for all bulk
Richardson numbers considered, that the time-averaged-kinetic-energy-maximising perturbations
are significantly suboptimal
at mixing compared to the 
mix-norm-minimising perturbations,
\tcr{and also that minimising the  mix-norm remains (for
  density-stratified
flows) a good
proxy for identifying perturbations which minimise the variance
at long times}. Although increasing stratification reduces the
mixing in general, mix-norm-minimising optimal perturbations can still
trigger substantial mixing for $Ri_b \lesssim 0.3$. By considering the
time evolution of the kinetic energy and potential energy reservoirs,
we find that such perturbations lead to a flow which, through Taylor
dispersion, very effectively converts perturbation kinetic energy into `available
potential energy', which in turn leads rapidly and irreversibly to
thorough and efficient mixing, with little energy
returned to the kinetic energy reservoirs.
\end{abstract}



\section{Introduction}

Irreversible mixing is ubiquitous in environmental and industrial
flows, and is a critical mechanism for ocean and atmosphere dynamics as
well as a crucial element in many manufacturing processes, occuring on
various length and time scales. Consequently, there has been 
great interest in understanding, quantifying and
finally improving the mixing properties of a vast range of flows,
whether to prevent or, alternatively to enhance, homogenisation,
notably through the triggering of  hydrodynamic instabilities. 
In particular recently,  maximization of the (transient)
perturbation kinetic energy, 
associated with the inherent non-normality of the linearized
Navier-Stokes operator, has been used as a convenient proxy for
optimising mixing efficiency without requiring a direct assesment of
well-mixedness,
which in a real sense is \tcr{both} an essentially nonlinear \tcr{and diffusive}
phenomenon, requiring as it does an (irreversible) modification of an
initial 
base or background scalar distribution. 

The fundamental mathematical hypothesis underlying this approach
(see e.g. \cite{AKB03,BAK05}) is 
that the `best' way (ultimately) to mix a scalar in a fluid flow  is
to  encourage flow instabilities or transiently growing perturbations,
presumably eventually triggering nonlinearities which would prevent
disturbances from fading away, 
thus ensuring scalar homogenization. 
On the other hand, recently developed mathematical tools, in
particular what we refer to here as the direct-adjoint-looping (DAL)
method (see \ref{sec:DAL} and references therein) have provided 
a new  algorithmic approach for inherently nonlinear optimisation of a
chosen quantity of interest, which can be an
appropriately chosen direct measure of mixing. 

Quantifying mixing is however not always an immediately
straightforward task, and the choice of a particular measure has the
potential to have implications for the optimal mixing strategy \tcr{within a finite time horizon}. Let us
consider $\rho({\bf x},t)$ to be a scalar field with diffusivity
$\kappa$ in a fluid with velocity field ${\bf U}({\bf x},t)$. 
If there are no sources or sinks in the domain, $\Omega$, of interest,  $\rho$ satisfies the advection-diffusion equation
\be
\label{adv-diff}
\fb{\pd \rho}{\pd t} \ + \ {\bf U}\cdot \nabla \rho \ =  \ \kappa \nabla^2 \rho.
\ee
Without loss of generality, we may assume that $\rho$ has spatial mean zero. Its variance
$\mathscr{V} = \| \rho \| ^2_2$,
where the \tcr{appropriately normalized} $L_2$-norm is 
defined as $\| X \|^2_2 =
\fb{1}{V_\Omega} \int_\Omega |X|^2 d\Omega$, provides a natural and
meaningful measure of the mixedness of the fluid as it quantifies the
deviation from the (zero) scalar spatial mean over the domain $\Omega$
\tcr{of volume $V_\Omega$}. The evolution equation for the variance is readily derived from (\ref{adv-diff}):
\be
\label{dVdt}
\fb{d}{dt} \| \rho \| ^2_2 \, = \, - 2\kappa \| \nabla \rho \| ^2_2,
\ee
indicating that the variance monotonically decreases toward zero, with
a decay rate determined exclusively 
by the diffusivity. From a mixing optimisation point of view, a first
issue arising from this equation  is that the variance decay rate depends only implicitly 
on the velocity field, through the velocity field inducing high
gradients in the scalar field. Indeed, efficient mixing occurs through
the (intermediate time) creation of strong local gradients, 
which are then ultimately smoothed by diffusion at small scale. 
A second practical issue is that (\ref{dVdt}) is not useful to describe
homogenisation in the theoretical limit of pure `stirring' i.e. as
$\kappa \rightarrow 0$.

Specifically to overcome this issue, 
\cite{MMP05} and \cite{DT06} have introduced and generalized what are
now commonly referred to as \textit{mix-norms}, namely Sobolev norms
of negative (possibly fractional) index (see \cite{T12} for 
a valuable review of mix-norms). 
These measures 
downplay the contribution of small scales by comparison to large
scales, 
and so scalar distributions with 
small values of a mix-norm can 
reasonably be thought to be (at least) approaching well-mixedness. 
Use of such mix-norms has already proved useful in a variety of mixing
optimisation problems where a passive scalar is advected by a pre-determined
velocity field, with or without sources (see for example \cite{Metal07}, \cite{TP08}, \cite{LTD11}).
Here, we will restrict our attention to the specific (particularly
computationally attractive) mix-norm for a zero spatial mean quantity $\rho$:
\be
\mathscr{M} = \| \nabla^{-1} \rho \|_2,\label{eq:mixnormdef}
\ee
where $\nabla^{-1} \rho$ is defined formally as $\nabla \Psi$, with
$\Psi$ the solution of the Poisson equation $\nabla^2 \Psi =
\rho$.
 Using (\ref{adv-diff}), it is possible to derive an evolution
 equation for the mix-norm which depends explicitly on both the
 diffusivity and (significantly) the flow velocity, 
\be
\fb{d }{dt} \| \nabla^{-1} \rho \|^2_2 \, = \,  \fb{2}{V_\Omega}\int_\Omega \nabla^{-1} \rho \cdot \nabla {\bf U} \cdot \nabla^{-1} \rho \, d\Omega \ - 2\kappa \, \| \rho \| ^2_2.\label{eq:mixevoleq}
\ee
Optimisation of the velocity field can therefore result in a faster
decrease of this mix-norm, compared to its decrease through pure
diffusion. Henceforth, we shall refer to this particular norm as {\itshape
  the} mix-norm. 

\tcr{It is important to remember
that other mix-norms (i.e. Sobolev norms with different negative indices) could be
considered instead. The choice of a particular index evidently
has the potential to impact
the identification of the optimal mixing strategy, by measuring how
`blurred' the scalar field appears in the light of a particular
mix-norm. In fact, the index can be thought as a way to  quantify the
typical level of \textit{filamentation} produced in the scalar field
by the target time considered for optimisation, a feature that is
essential in the mixing process and (to some extent) related to the
criterion proposed by \cite{kuka09} to quantify mixing based on
exposure (a function of the interface area and sharpness). 
The `optimal' choice of an optimisation norm
(in the sense of the choice of the value of the negative
index of a particular Sobolev norm)
 is a non-trivial matter, depending potentially on the optimisation time horizon,
the actual problem of interest, and even what measure is used to
identify the  `optimal
choice'. Indeed, \cite{MMP05} actually argued, from the viewpoint of
ergodic theory, that the most `natural' choice of negative 
index is $-1/2$ in such Sobolev norms.  However, there is some preliminary 
evidence (see \cite{vermach_thesis} for further details)
that the `mix-norm' (as defined in (\ref{eq:mixnormdef}) with index $-1$) is 
a solid general-purpose choice, and is also computationally convenient
because of its simple structure in spectral space.}

\tcr{Investigation of the `optimal choice' of the index is 
beyond the scope of this paper, not least because 
we wish to compare with previous studies which have used the mix-norm
as defined in (\ref{eq:mixnormdef}). In particular, using the DAL method, and this particular choice
  of the index in the mix-norm,}  
Foures, Caulfield \& Schmid (2014) considered the time-evolution of
the mixing 
of a passive scalar, subject to an
`optimal' initial velocity perturbation of finite energy, where the
subsequent freely evolving incompressible velocity field is the 
solution of the fully nonlinear (yet two-dimensional) Navier-Stokes
equations
in a plane channel flow 
driven by a constant pressure gradient. The initial 
scalar distribution was essentially two layer, with a thin 
interface where the initial concentration changed smoothly 
between the two layer values.
They considered three different 
optimisation problems: 
 maximisation of the time-averaged perturbation kinetic energy and
 minimisation of both the  variance and the mix-norm (defined above)
 at a range of target times. 

They made three key observations. First,
 the perturbations which optimised perturbation growth were
 significantly less effective at mixing the scalar than the
 perturbations which minimised either the variance (the natural
 measure of mixedness) or the mix-norm at the target time. 
Second, minimising the mix-norm proved to be a good proxy for 
minimising the variance over long-time horizons. Indeed,  
particularly conveniently computationally, minimising the mix-norm
over short time horizons proved to yield
a better approximation to the 
initial perturbation that minimised variance over a long time than
minimising the variance over the same short time horizons.
Third, the actual mixing process induced
by such optimal perturbations could be categorised 
as a three-stage process: `transport'; then `dispersion'; then `relaxation'.
The transport induced by the (optimal) initial perturbations
induced the scalar distribution to 
take the qualitative structure of alternating vertical
structures. These alternating `stripes' are then distorted into
`chevrons' and then dispersed
(principally by the mean shear) via so-called Taylor
dispersion \citep{Taylor53}, thus homogenising the scalar field until
the onset of the final stage of purely diffusive relaxation to a
completely mixing state.

Building in particular on this study, our aims herein are twofold.
First, we aim to extend this study to the situation where the scalar
$\rho$ is \emph{active} (i.e. $\rho$ is the fluid density in a
gravitational field),
and so  buoyancy forces (can) play a central role in the flow
evolution.
In particular, we investigate the hypothesis that the first two key observations 
mentioned above also apply in density-stratified fluids.
\tcr{Therefore,}
 we aim
to demonstrate
 both the inadequacy of maximising perturbation kinetic energy  and the
 usefulness  of minimising \tcr{(this particular)} mix-norm for identifying optimal initial
 perturbations 
to induce effective mixing strategies in stratified flows with finite
diffusivity. 
Second, we wish to gain physical 
insight into 
the effect of buoyancy forces
on the identified
time-evolution of such optimal initial perturbations. Specifically, we wish
 to understand whether mixing can still 
  occur even when density stratification acts to
suppress the non-trivial vertical velocities of denser and lighter
fluid required by the `transport' stage of the flow evolution
described
above which is central to `optimal' mixing of  a passive scalar.

To achieve these two aims, the rest of this paper is organised as
follows.
In section 2, we present our numerical model, and define precisely the two
optimisation 
problems which we consider \tcr{in detail}, based around maximisation 
of time-averaged perturbation kinetic energy, and minimisation 
of the mix-norm for our scalar field at the target time.  In section 3, we describe the physical
structures which develop during the flow evolution associated with the
various `optimal' perturbations, focussing in particular on
the dependence of this flow evolution on increasing stratification. 
\tcr{We also demonstrate that mix-norm minimisation over relatively
  short 
target times continues to be a good proxy for variance minimisation
over
long target times for flows with active scalars, confirming and
generalizing 
the results of \cite{FCS14}.}
In section 4, we consider quantitatively the flow energetics of 
typical
cases of both optimisation problems in the presence of non-trivial
stratification, specifically to gain insight into the differing
(irreversible) mixing properties of the two flows. 
Finally, we draw our conclusions in section 5, and discuss possible
future avenues of research.


\section{Optimal mixing problem formulation}


\subsection{Flow configuration and governing equations}

We wish to determine the (nonlinear) initial velocity perturbation
$\bf u_0$ of 
fixed 
kinetic energy  to a density-stratified, plane Poiseuille
two-dimensional flow which  optimises `mixing' (defined in two
different mathematical ways)
 of the fluid over a chosen time horizon.
 The (dimensionless) pressure-driven Poiseuille background base flow ${\bf
   U}=U(y) {\bf e_x}$, where 
\be
U(y) \ = 1- y^2 ,\label{eq:baseu}
\ee
is prescribed in a periodic channel of dimensionless length $L_x=4\pi$
and width $L_y=2$.
Lengths have thus been scaled with the (dimensional) channel half-depth $h^*$ and
the 
maximum (base) flow speed $U_m^*$. This length of channel has been
chosen to allow for the possibility of a wide range of streamwise
scales, while still being computationally inexpensive. Undoubtedly, there
is an issue with the periodic boundary conditions
leading to re-entrant flow of partially-mixed fluid advected
principally by the background base flow, particularly near the middle
of the channel. However, it is important to appreciate that here we
are 
principally focussed on a proof-of-concept approach to demonstrate
that the DAL method is useful to `optimise' mixing flows, 
rather than an exhaustive parameter study of more
realistic stratified mixing in pressure-driven channel flows.

We denote the dimensional density  difference across the channel as $ \rho^*_0$ and the 
mean density as $\bar{\rho}^* \gg \rho_0^*$, 
so that we may make the Boussinesq approximation, 
and we also implicitly consider a fluid with a linear
equation of state.
For mathematical convenience, it is natural 
to consider zero-mean quantities, and so we consider the density
deviation field $\rho^*$
from the mean $\bar{\rho}^*$, 
i.e.
\be
\rho^* ({\bf{x^*}},t^*) = \rho_T^* - \bar{\rho}^*,
\ee
where $\rho_T^*$ is the total density field. Scaling the deviation density field with $\rho^*_0$, 
we choose the initial (dimensionless) density deviation field
$\rho_i({\bf x})$ to be
\be
\rho_i({\bf x})=
\rho({\bf x},0) \ = - \tfrac{1}{2} \, \text{erf} \left(\fb{y}{\delta_0}\right),\label{eq:rhotzero}
\ee
so that we consider a 
stably-stratified miscible fluid initially arranged in two layers
separated by
a diffusive interface of typical thickness $\delta_0 = 0.025$. The
dimensionless, nonlinear Boussinesq system (with an implicit linear equation of
state) governing the evolution of
the perturbation velocity ${\bf u}(x,y,t)$, perturbation pressure
$p(x,y,t) $ and density deviation $\rho(x,y,t)$ then depends on three
dimensionless parameters:
\begin{align}
\label{NS}
\fb{\pd \bf u}{\pd t} \ + \ \big(\bf u + U\big)\cdot \nabla \big({\bf u + U}\big) \ &= \ - \  \nabla p \ -\  Ri_b \,\rho {\bf e_y} \ + \ \Rey^{-1} \nabla^2 {\bf u},\\
\label{drhodt}
\fb{\pd \rho}{\pd t} \ + \ \big({\bf u + U}\big)\cdot \nabla \rho \ &=  \ \Pen^{-1} \nabla^2 \rho,\\
\label{divu}
\nabla \cdot {\bf u} \ &= \ 0,
\end{align}
namely the Reynolds number $\Rey$, the P\'eclet number $\Pen$ and the bulk
Richardson number $Ri_b$,
defined as\bme
\be\te
\Rey=\fb{U_m^* h^*}{\nu^*}, \qquad \Pen=\fb{U_m^*h^*}{\kappa^*}= \Rey
\frac{\nu^*}{\kappa^*}=\Rey \Pran, \qquad Ri_b=\fb{g^* \rho_0^* h^*}{ \bar{\rho}^* U_m^{*2}}.
\ee
\eme
In these expressions,
 $\nu^*$ is
the  kinematic viscosity of the fluid, $\kappa^*$ is the  thermal
diffusivity of the fluid, $\Pran$ is the Prandtl number and $g^*$ is the
gravitational acceleration.  We choose $\Pran=1$ and $\Rey=500$
throughout this study for ease of comparison with \citet{FCS14}, but
vary $Ri_b$ to investigate the extent to which buoyancy effects
modify the flow evolution.
Periodicity is assumed in the streamwise direction while
we impose no-slip, no-flux boundary conditions
at the channel walls at $y =\pm 1$, and so
\bme
\label{BCdir}
\be\te
{\bf u = 0 }, \qquad \pd_y \rho=0, \qquad \text{and} \quad \pd_y p=0.
\ee
\eme


\subsection{Variational problem}


\subsubsection{Choice of cost functional}

Optimisation requires defining a quantity of interest (the objective
functional) to be extremised subject to a set of constraints.
 As discussed in the Introduction,  
one of the primary aims of this study is to extend the results of
\cite{FCS14}  to consider mixing of an `active' scalar, where buoyancy effects play
a role. In particular, the results of \cite{FCS14} (see their figure
7, in particular 7f and 7i) suggest that, whereas mix-norm and
variance minimisation over long target times yield similar values of
the terminal variance when the initial, optimal perturbation flows
freely evolve in time, mix-norm minimisation over short target times
is likely to achieve significantly smaller long-term variance than
variance-optimisation. In that respect, mix-norm minimisation can then
be viewed as a proxy for long-term minimisation of the scalar variance
even for  relatively short target times, thus
allowing for considerable computational savings. In what follows the
scalar mix-norm is therefore the quantity to be minimised at target
time $T$ while we assess the well-mixedness of the optimal flow using
the variance at later times as a diagnostic variable. 
\tcr{As discussed in the Introduction, although other Sobolev
norms could possibly be chosen, we are particularly interested
in determining whether the specific 
observations of \cite{FCS14} (concerning the usefulness
of  mix-norm-minimisation
calculations over intermediate
target times  as a proxy for variance-minimisation at long 
target times) generalise to 
density-stratified flows. Therefore, we focus here on the same 
mix-norm and (typically) a single choice of intermediate target time horizon.
}

We are also interested in investigating 
in a density-stratified flow whether perturbations
which maximise the time-averaged perturbation kinetic
energy are
still significantly worse at mixing than
mix-norm minimising perturbations. Therefore,  
the objective functional $\mathscr{J}$ we consider is  defined as
\be
\mathscr{J}\Big({\bf u}, \rho \Big) = \fb{1-\alpha}{2T} \int_0^T \| {\bf u}({\bf x},t) \| ^2_2 \, dt \  + \ \fb{\alpha}{2} \| \nabla^{-\beta} \rho({\bf x},T) \| ^2_2.
\ee
\tcr{Here, the parameters $\alpha$ and $\beta$ are switches  which can take only the values
$\alpha = 0$ (when the value of $\beta$ is irrelevant) or $\alpha = 1$
with $\beta=0$ or $\beta =1$, and are set depending on the quantity we wish to extremise.} 
Setting $\alpha=0$ yields the objective functional appropriate for the
identification of a  
perturbation which maximises  the time-averaged perturbation kinetic energy 
developed by the perturbation velocity field throughout the time
interval
 $[0, T]$. On the other hand, setting $\alpha=1=\beta$ yields the objective functional appropriate for the
identification of a  
perturbation which minimises the mix-norm at time $T$, 
\tcr{while  setting $\alpha=1$, $\beta=0$
yields the objective functional appropriate for the
identification of a  
perturbation which minimises the variance at time $T$.}

 We wish to
extremise this objective  functional over all possible choices of initial
velocity perturbation ${\bf u}({\bf x},0)={\bf u_0}$ subject to the
constraints that the governing (forward or `direct') equations
(\ref{NS})-(\ref{divu}) are satisfied at all points in space and time
by the perturbation velocity field ${\bf u}$, the perturbation
pressure $p$, and the deviation density $\rho$.
 The augmented Lagrangian is therefore
\[\mathscr{L}\Big({\bf u}, p, \rho, {\bf u_0}, \tilde{\bf u}, \tilde{p}, \tilde{\rho}, \tilde{\bf u}_{\bf 0}\Big) \ = \ \mathscr{J}\Big({\bf u}, \rho\Big)   
\ - \   \lel \vt \ , \fb{\pd {\bf u}}{\pd t} + {\bf N(u)} + \nabla p + Ri_b \, \rho {\bf e_y} - Re^{-1} \nabla^2 {\bf u} \rir \]
\be 
\label{Ldef}
\qquad \qquad \text{...}  \quad - \lel \rt \ , \fb{\pd \rho}{\pd t} + \big({\bf u + U}\big)\cdot \nabla \rho - Pe^{-1} \nabla^2 \rho \rir 
\ee
\[ \qquad \qquad \text{...}  \quad - \lel \tilde{p} \ , \nabla \cdot {\bf u} \rir \ - \ \bigg[\tilde{\bf u}_{\bf 0} \ , {\bf u}({\bf x},0) - {\bf u_0} \bigg],\]
\noindent where we have introduced the nonlinear advection operator
${\bf N(u)}= \big({\bf u + U}\big)\cdot \nabla \big({\bf u + U}\big)$
and the Lagrange multipliers $\tilde{\bf u}$, $\tilde{\rho}$,
$\tilde{p}$ and $\tilde{\bf u}_{\bf 0}$ (which imposes the initial condition). The two inner products
are defined as 
\begin{subequations}
\be
\lel {\bf u,v}\rir = \fb{1}{V_\Omega T} \int_0^T  \int_\Omega  {\bf u \cdot v} \, d\Omega dt,
\ee
\be
\text{and} \qquad \big[{\bf u,v}\big] = \fb{1}{V_\Omega} \int_\Omega
{\bf u \cdot v} \, d\Omega, 
\ee
\end{subequations}
where $V_\Omega=L_x L_y = 8 \pi$ is here the flow `volume', i.e. the area
of the computational domain. It is important to appreciate that the steady
background base velocity distribution $U(y)$ as defined in
 (\ref{eq:baseu}) is imposed by a constant pressure gradient which
 does not enter into these equations. However, there is no assumption 
that the perturbation velocity ${\bf u}(x,y,t)$
is small compared to $U$, and indeed it 
is entirely possible that at least transiently the
horizontally-averaged
streamwise velocity may vary non-trivially from $U(y)$, due
to the inherently nonlinear nature of the flow perturbations.


\subsubsection{Formulation of the Direct-Adjoint-Looping (DAL) problem}
\label{sec:DAL}

Setting the first variations of  (\ref{Ldef})
with respect to the Lagrange multipliers to zero naturally recovers the imposed
constraints, while setting to zero the first variations with respect
to the direct flow variables ${\bf u}$, $\rho$ and $p$ yield evolution
equations for the adjoint variables (or sensitivities, see \cite{H95}
for more description):
\begin{align}
\label{adj1}
\fb{\pd \vt}{\pd \tau} \ + {\bf \tilde{N}}(\vt)  & \ = \  \nabla \tilde{p} \ + Re^{-1} \nabla^2 \vt  - \rt \ \nabla \rho \ + (1-\alpha) {\bf u},\\
\label{adj2}
\fb{\pd \rt}{\pd \tau} \ - \big({\bf u + U}\big)\cdot \nabla \rt & \ = \ - Ri_b \ \vt \cdot {\bf e_y} \ + Pe^{-1} \nabla^2 \rt,\\
\label{adj3}
\nabla \cdot \vt  & \ = \ 0,
\end{align}
with the adjoint operator ${\bf \tilde{N}(v)} = v_j
\pd_i \big(u+U\big)_j {\bf e_i} - \big({\bf u+U}\big) \cdot \nabla
{\bf v}$ using Einstein summation notation. 
\tcr{In these expressions, ${\bf e_y}$ and ${\bf e_i}$ denote
unit vectors in the $y-$direction and the ${\rm i^{th}}$ direction
respectively.}
Note that, as is
conventional, the adjoint equations (\ref{adj1})-(\ref{adj3}) are
well-posed when integrated \emph{backwards} in time $\tau=T-t$ from
the terminal time $T$ to the initial time $0$, as is apparent from the
relative signs of the time derivative and diffusive terms.
Interestingly, 
the velocity field's dynamical dependence on the density $\rho$
through the buoyancy term results in a sensitivity-dependent forcing
term $- Ri_b \ \vt \cdot {\bf e_y}$ in the (backward) evolution
equation for the adjoint density.

Of course, 
various boundary (in both space and time) terms appear when constructing  the adjoint equations
(\ref{adj1})-(\ref{adj3}).
Some naturally vanish due to the imposed boundary conditions on both the direct velocity and density fields (\ref{BCdir}). 
The requirement that the remaining boundary terms vanish  provide terminal conditions for the adjoint velocity and scalar fields at $\tau=0$:
\bme
\label{termC}
\be
{\bf \tilde{u}}={\bf 0},  \qquad \qquad \tilde{\rho}=(-1)^{\beta}
\alpha \nabla^{-2 \beta} \rho,
\ee
\eme
as well as an  `initial' or compatibility condition for the adjoint velocity field at $\tau=T$:
\be
{\bf \tilde{u}}={\bf \tilde{u}_0},
\ee
and natural  boundary conditions for the sensitivities at the channel walls ($y = \pm 1$):
\bme
\be\te
{\bf \tilde{u} \cdot n} = 0, \qquad {\bf \tilde{u} \times n = 0}, \qquad \text{and} \qquad 
\pd_y \tilde{\rho}=0.
\ee
Periodicity is also assumed in the streamwise direction for the
adjoint variables. There is no implied choice of a boundary condition
for the adjoint pressure field.  Symmetry considerations suggest however our choice of the same Neumann condition as in the direct problem (\ref{BCdir}):
\be\se
\pd_y \tilde{p}=0 \qquad \qquad \text{for} \quad y=\pm 1.
\ee
\eme

Finally, 
the gradient of $\mathscr{L}$ with respect to the choice of the
initial velocity perturbation  $\bf u_0$ yields
\be
\nabla_{\bf u_0} \mathscr{L}= {\bf \tilde{u}_0}. \label{eq:u0grad}
\ee
As discussed in more detail in Foures, Caulfield \& Schmid (2013) and
\citet{FCS14}, 
due to numerical convergence requirements the normalisation of the
initial velocity perturbation to have a fixed energy $K_0$ is not enforced by means
of an additional Lagrange multiplier in the definition of the
augmented Lagrangian.
 Instead, we use the same approach as  the one proposed by
 \cite{FCS13},
 and geometrically enforce the normalisation constraint directly
 within the optimisation routine. 
We require
\be
K_0 = \tfrac{1}{2}\big \| {\bf u_0}\| ^2_2=0.01, 
\label{eq:norm}
\ee
a value chosen
consistently
with previous work to  avoid scale-separation between the perturbation and the base flow 
velocities and also to ensure that the nonlinear terms in (\ref{NS})
play a non-negligible 
role in the flow evolution. It is important to stress again 
that a wider parameter study would be valuable, and that we are
here principally concerned with a proof-of-concept demonstrating that
the use of the DAL method has the potential 
to yield valuable insight into `optimal' mixing problems
\tcr{in stratified flows, where the density field is playing an `active'
  or dynamic
  role}.

We then calculate (see \cite{FCS13} for further details) 
the component of the gradient (\ref{eq:u0grad}) on the hypersurface
which satisfies the normalization condition (\ref{eq:norm}). We 
 denote this component by $\nabla_{\bf u_0} \mathscr{L}^{\perp}$.
Formally, convergence has occured when $\nabla_{\bf u_0}
\mathscr{L}^{\perp}=0$. This implies that 
the evolution of the adjoint equations has yielded ${\bf \tilde{ u}_0}
\propto {\bf u_0}$, and so the only
way in which the Lagrangian can be increased or decreased is by moving
parallel (or anti-parallel) to the present choice of $\bf{u}_0$, 
thus inevitably violating the normalisation constraint.

In practice the DAL method iteratively computes the optimal initial
perturbation ${\bf u_0}$ as follows.
 A first guess, associated with the given kinetic energy density 
 budget $K_0$,  is input as an initial condition to the direct, fully
 nonlinear Navier Stokes equations (\ref{NS})-(\ref{divu})
solver and integrated forward in time up to the target time $T$. 
As the adjoint evolution 
equations (\ref{adj1}) and (\ref{adj2})  depend upon the direct
variables $\bf u$ and $\rho$, they are stored at intermediate
`checkpoint' times.  
The terminal conditions (\ref{termC}) then provide a starting point for the
backward time-integration of the adjoint problem
(\ref{adj1})-(\ref{adj3}), 
with further ancillary forward integrations
from each of the checkpoints being used (along with fractional step
interpolations when needed) to construct direct 
variables at each of the required
time instants for the adjoint integration back to the initial
time $t=0$.  The initial adjoint velocity field $\bf \tilde{u}_0$ then
provides useful gradient information in
(\ref{eq:u0grad}),
through projection of $\bf \tilde{u}_0$ onto the
hypersurface
which satisfies the normalisation constraint, thus constructing the
appropriate component of the gradient
$\nabla_{\bf u_0} \mathscr{L}^{\perp}$. This projected gradient is
used in a conjugate gradient algorithm to update the initial condition
`guess'
for ${\bf u_0}$,
and the 
 looping process is iterated until numerical
convergence is reached, which is defined operationally in terms of the appropriate
scaled residual 
\be
r = \fb{\| \nabla_{\bf u_0} \mathscr{L}^{\perp} \| ^2_2}{\| \nabla_{\bf
    u_0} \mathscr{L} \| ^2_2} \leq 10^{-4},
\ee
reaching a sufficiently small threshold.
A review of this method has been presented for 
instance in \cite{Peter07}, as well as in \cite{K014} and \cite{LB14}
in the context of stability analysis with fully nonlinear dynamics. 


\subsection{Numerical model}
All the results presented here have been produced with a modified
version of the DAL method code wrapping the two-dimensional 
 Navier-Stokes equations solver  developed by \cite{FCS13} (see also
 \cite{FCS14}).
The key \tcr{(and dynamically significant)} modification is 
that the scalar field
has been made active, through the addition 
of the appropriate buoyancy terms.  The code uses second-order finite
differences on a staggered grid in the $y$ (spanwise) direction and
spectral decomposition in the $x$ (streamwise)
direction. Divergence-free conditions for the direct and adjoint
velocity fields are ensured by means of a Chorin's projection
algorithm \citep{C67,C68}. Time integration of diffusive terms is
dealt with using a Crank-Nicolson scheme. A Runge-Kutta fourth order
scheme is used for the remaining terms except for the forcing term
$(1-\alpha) {\bf u}$ in (\ref{adj1}), which is treated implicitly. The
simulations presented here have all been performed with $300$ Fourier
modes in the $x$ direction and $200$ gridpoints in the $y$ direction,
a resolution which 
convergence tests proved to be sufficient.


\section{Numerical results}

As noted in the Introduction, all  the simulations presented here have been performed at Reynolds
number $Re=500$ and P\'eclet number $Pe=500$ with the bulk
Richardson number $0.01 \leq Ri_b \leq 1.0$. 
We also validated our simulations
by reproducing the 
passive scalar mixing ($Ri_b=0$) 
results of \cite{FCS14}. 
\tcr{In general, we have} considered both time-averaged-kinetic-energy
maximisation ($\alpha=0$) and mix-norm minimisation
($\alpha=1=\beta$), with a target time $T=10$. 
\tcr{We have also considered variance minimisation ($\alpha=1$,
  $\beta=0$), with the principal objective \tcr{being} to investigate
whether mix-norm minimisation
remains a good proxy for variance minimisation in density-stratified flows.}
Reaching numerical convergence is particularly challenging here
because the problem, although two-dimensional, is fully nonlinear, and
furthermore the adjoint evolution equations (\ref{adj1})-(\ref{adj2})
are forced by the direct variables - all the more so for the $\alpha=0$ case, where (\ref{adj1}) is driven by an additional forcing term depending on the direct velocity field. This last feature requires very accurate evaluation of the direct fields for the sensitivities to be in turn correctly computed. 
It should also be  emphasised that the DAL method only identifies
local extrema. The first `guess'  initial condition  consists of
Gaussian noise in all the cases considered, thus hopefully ensuring that the
initial perturbation is unlikely to have zero projection onto the true
`optimal' initial perturbation. \tcr{In addition, the optimisation
  procedure was systematically initiated with different initial
  perturbations and always resulted in identifying the same (local)
  extremum.
\tcr{This observation is reassuring, although}  it cannot be ensured
that this extremum 
\tcr{actually} is a global \tcr{extremum} due to the non-convexity of the \tcr{optimisation} problem.}


\subsection{Optimal initial perturbations in  stratified fluids}\label{sec:opt_strat}

\begin{figure}
\begin{center}
\includegraphics[width=\textwidth]{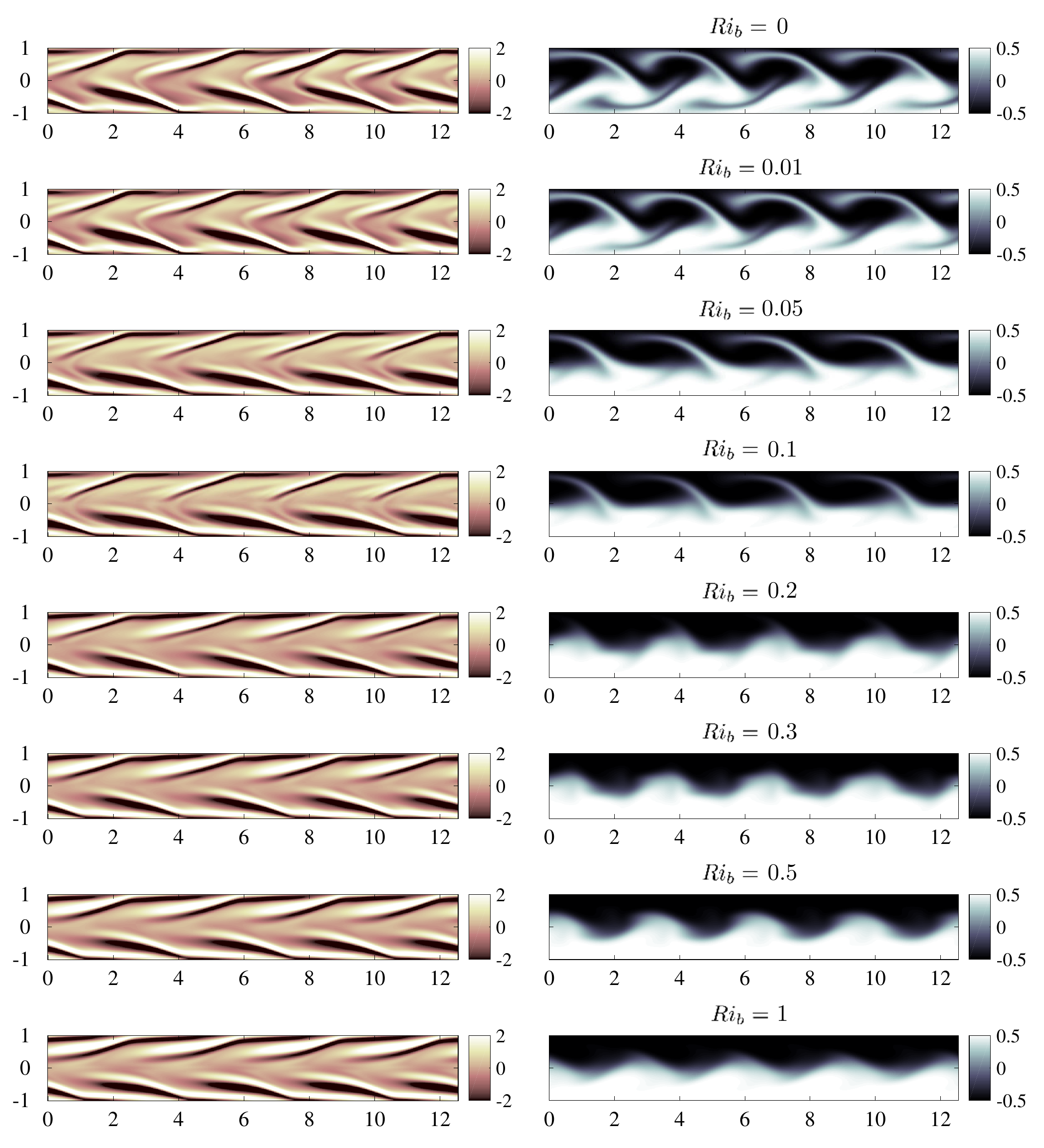}
\captionsetup{width=0.9\textwidth}
\caption{\small{Initial perturbation vorticity $\bnabla \times {\bf u_0}$ (left) and terminal
    deviation density  field $\rho$ at T=10 (right)  for
    the time-averaged-kinetic-energy-maximisation problem for bulk Richardson number $Ri_b=$
    (top to bottom): $0$; $0.05$; 
$0.1$; $0.15$; $0.2$; $0.3$; $0.5$; $1$.     
Movies showing the time evolution of these fields are available as 
supplementary materials.   
}}
\label{fig:rhoE}
\end{center}
\end{figure}

\begin{figure}
\begin{center}
\includegraphics[width=\textwidth]{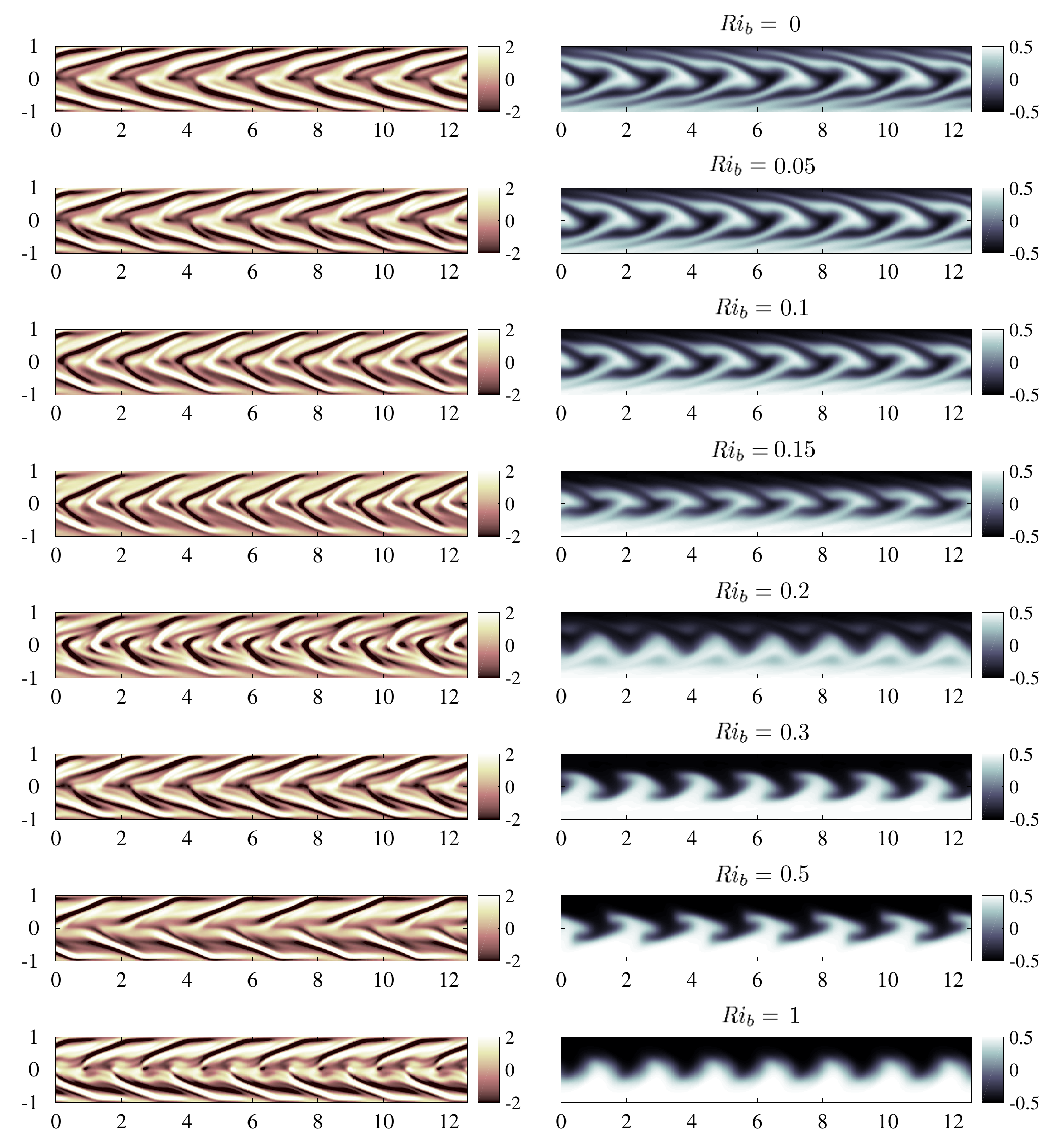}
\captionsetup{width=0.9\textwidth}
\caption{\small{Initial perturbation vorticity $\bnabla \times {\bf u_0}$ (left) and terminal
    deviation density  field $\rho$ at T=10 (right) for
    the mix-norm-minimising problem  for bulk Richardson number
    $Ri_b=$
(top to bottom): $0$; $0.01$; $0.05$; 
$0.1$; $0.2$; $0.3$; $0.5$; $1$.    
Movies showing the time evolution of these fields are available as 
supplementary materials.}}
\label{fig:rhoM}
\end{center}
\end{figure}

We plot the optimal initial perturbations 
associated with the time-averaged-kinetic-energy-maximisation problem 
and the mix-norm-minimising problem
in the left columns of figures \ref{fig:rhoE} and \ref{fig:rhoM} respectively, in terms of the (initial) vorticity field ${\boldsymbol
   \omega_{\mathbf 0} }= \bnabla \times \bf u_0$ for flows with eight
   different values of the bulk Richardson number $Ri_b$. We plot the
   associated terminal scalar field $\rho(x,y,T)$ in the right column of figures \ref{fig:rhoE} and \ref{fig:rhoM}. 
Consistently with the observations of 
\cite{FCS14} for equivalent optimisation problems
for a  passive scalar with the same optimisation horizon $T=10$ , the
dominant wavenumber selected for the optimal perturbation field is
$m=4$ for the time-averaged-kinetic-energy-maximisation problem and $m=7$ for
the mix-norm minimisation problem.  
(\cite{FCS14} found these optimal wavenumbers decrease as the
time-horizon $T$ increases for passive scalar flows). 
For both problems when $Ri_b=0$,  the initial vorticity field consists of inclined,
stretched vortices tilted into  the shear of the base flow $U(y)$, as
defined in (\ref{eq:baseu}). 

This flow configuration naturally favours transient energy growth due
to the so-called Orr mechanism \citep{Orr07}. The shear tilts the
initially elongated vortices, transiently
reducing their aspect ratio  and their perimeters and hence increasing
their energy by (close-to) conservation of circulation, before
again stretching them out into elongated vortices now inclined
with the shear. 
Importantly, these optimal perturbation fields also take advantage of
Taylor dispersion \citep{Taylor53} to homogenise the density field. 
As part of the 
`transport' phase as described in the Introduction
and in \citet{FCS14}, 
 the perturbations transport dense fluid 
upwards and light fluid downwards
towards the channel walls, where the locally strong base shear flow
near the wall inevitably both sharpens the density gradients and
further encourages filamentation and thus enhances diffusion. 

However, the behaviour as stratification increases is markedly different for
the two optimisation problems. 
As $Ri_b$ increases to non-trivial values between $0.05$ and $1$,  the observed patterns (both for initial
vorticity, and scalar distribution at terminal time)
for the  mix-norm minimisation problem  do not change qualitatively. 
Increasing stratification does tend to inhibit transport of the
stretched interface toward the walls, eventually resulting in the
mixing being ineffective at sufficiently high $Ri_b$. As can be seen in
figure \ref{fig:rhoM} showing the terminal density field at $Ri_b=1$, 
the characteristic protruding, chevron-like structures 
which develop  at lower $Ri_b$ have almost disappeared. Essentially,
the  interface is merely slightly displaced upwards  and downwards,
and so Taylor
dispersion does not take place and the 
only irreversible
modification
is associated with relatively weak diffusion 
at the interface.

This transition 
to essentially ineffective
mixing occurs in flows with much smaller $Ri_b$ 
for the time-averaged-kinetic-energy-maximisation problems.
As plotted  in figure \ref{fig:rhoE}, the two layers appear to remain
completely distinct for a Richardson number as low as $Ri_b=0.2$ for
these problems. 
Interestingly, stratification appears to affect the evolution of 
the density field for the time-averaged-kinetic-energy-maximisation 
problem in a more subtle way than it does for the mix-norm
minimisation problem. 
Indeed, the complicated density  structure  observed for flows with
low $Ri_b$ results from 
strong  filamentation and subsequent roll-up of the already-folded
interface, which starts to develop as soon as the peak in kinetic energy
is reached.
In the absence of
buoyancy forces ($Ri_b=0$) the resulting structure essentially exhibits
up-down symmetry, a symmetry which appears to be  broken for the flows
with relatively weak stratification (i.e. with
$Ri_b=0.01-0.1$). Furthermore, this apparent symmetry breaking is
accompanied by a progressive disappearance of the filaments as $Ri_b$
increases towards $Ri_b=0.2$, where they actually  vanish. 
However, looking at the time evolution of the density field 
(as in the accompanying supplementary material movies)  reveals that 
rolling up of the  density distribution occurs both in the upper
half-plane and the lower half-plane, alternating in time. 
This particular process of alternating roll-up is 
a dynamical process  which is switched off for flows with  $Ri_b
\gtrsim 0.2$. 

Finally,  
it is important to appreciate  that the apparently anomalous behaviour
of the mix-norm-minimising problem for  $Ri_b=0.2$ (in point of fact
very similar to the behaviour of a flow with  $Ri_b=0.25$, which for
reasons of space is not displayed \tcr{in figure} \ref{fig:rhoM}),  is best
understood as a somewhat misleading snapshot in its time evolution.
 The  transition in behaviour  as $Ri_b$ increases from $Ri_b=0.15$ to
 $Ri_b=0.3$ occurs as the largest parts of the periodic density
 `chevrons' fall back toward the centre of the channel under the
 increasing effect of gravity earlier and earlier in the flow
 evolution,
as is apparent from the movies of the flow evolution available as
supplementary materials.
Indeed their filamented remnants are  still visible near the
boundaries in the flow for  $Ri_b=0.2$, 
and the snapshot shown in the figure is just at the `wrong' stage in
the flow
evolution to show the chevron structures clearly.  As the
stratification
gets stronger, with  $Ri_b \gtrsim 0.3$, 
buoyancy forces
suppress
the  chevron
structure
developing so that 
the filaments cease
 to interact significantly
with the heightened shear near the 
channel walls.


\subsection{Comparing the objective functionals}

Such qualitative comparisons of the terminal density fields 
for various $Ri_b$ are highly suggestive that the 
time-averaged-kinetic-energy-maximising 
initial perturbation is less effective
than the mix-norm minimising initial perturbation
at stratified mixing at $Re=Pe=500$, 
at least for $Ri_b > 0.2$. This suggestion can be confirmed and
extended 
to all Richardson numbers by quantitatively assessing the
well-mixedness of the flow at long times. 

Even though the mix-norm is used in the definition of the cost
functional in our optimisation problems  (for the computational reasons
given above) 
the variance is still the natural  physical measure which should be
used as a diagnostic for the well-mixedness or homogeneisation of the
initially stratified
fluid.
One way to consider an aspect of 
the time-dependent `quality' of mixing by the various flows is to
compute an instantaneous  normalised variance (following \cite{FCS14}) 
$V(t)$, defined as the ratio
of the variance of the density field at instant $t$ to the variance
of the initial base  density field at the same instant under the 
action of pure diffusion:
\be
\label{defV}
V(t) = \fb{\| \rho(x,y,t) \|^2_2}{ \| \rho_{d}(y,t)\| ^2_2},
\ee
where $\rho_d(y,t)$ is the one-dimensional solution of
the purely diffusion version of the density deviation evolution
equation (\ref{drhodt}) 
(assuming $\bf u = \bf U = 0$) with initial condition  $\rho_i({\bf
  {x}})$ 
as given in (\ref{eq:rhotzero}). Small values of $V(t)$
therefore denote well-mixedness of the flow induced by fluid motions.

\begin{figure}
\begin{center}
\includegraphics[width=0.8\textwidth,center,trim=0 40 0 0,clip=true]{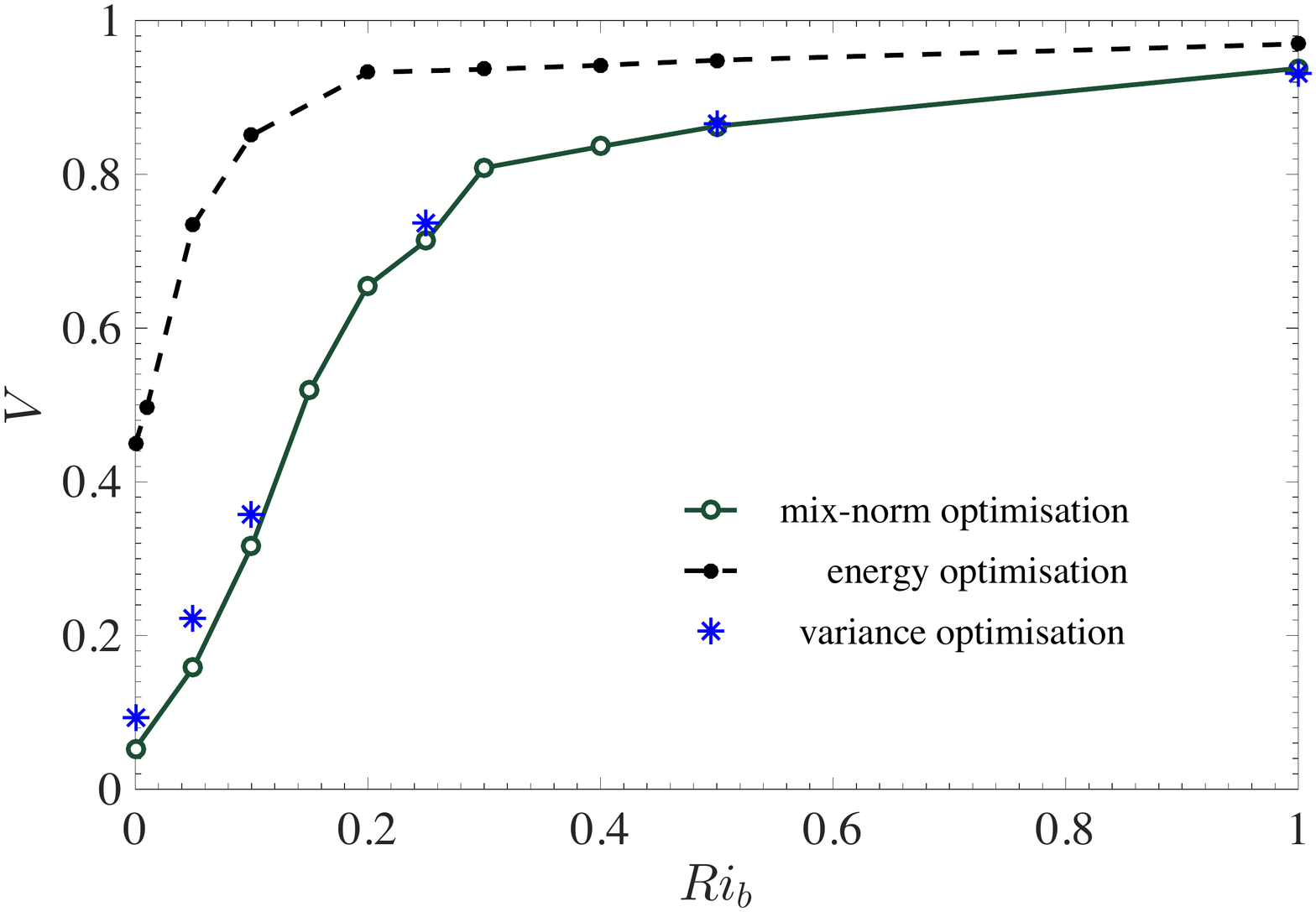}
\captionsetup{width=0.8\textwidth}
\caption{\small{Variation with $Ri_b$ of the scaled variance
$V$  (as defined in (\ref{defV})) at time $t=30$   for:
time-averaged-kinetic-energy-maximisation problem flows (plotted with a
dashed line) and mix-norm minimisation problem flows (solid line), all
for the intermediate time horizon $T=10$. \tcr{For comparison, stars mark the values
  of 
the scaled variance at $t=30$ for variance-minimisation problem flows, 
calculated for the same intermediate time horizon of $T=10$.}
}
} 
\label{fig:VvsRi}
\end{center}
\end{figure}

\begin{figure}
\begin{center}
\begin{overpic}[height=0.32\textwidth,trim=5 10 0 0,clip=true]{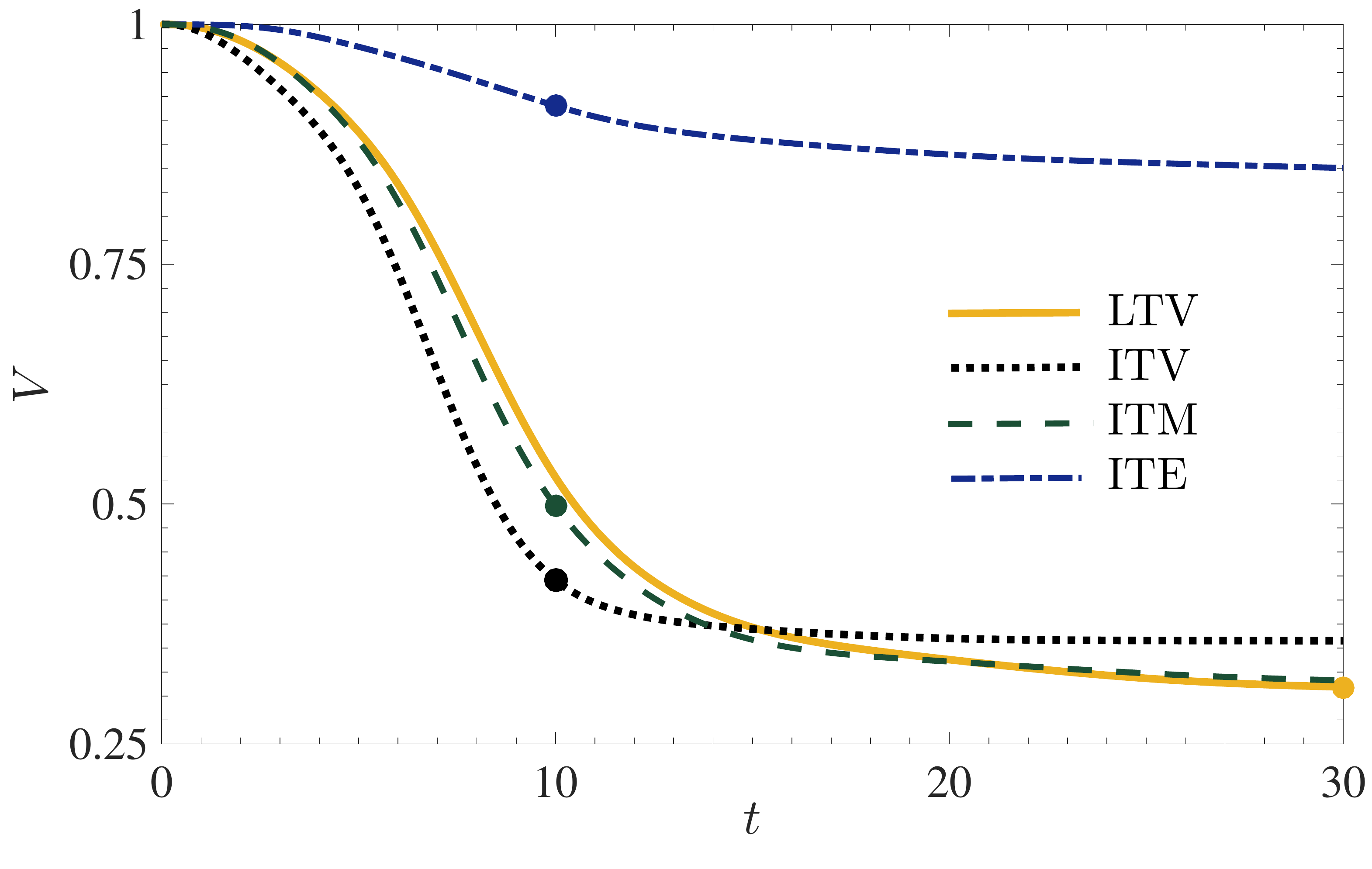}
\put(85,55){\colorbox{white}{\parbox{0.02\linewidth}{a)}}}
\end{overpic}
\begin{overpic}[height=0.32\textwidth]{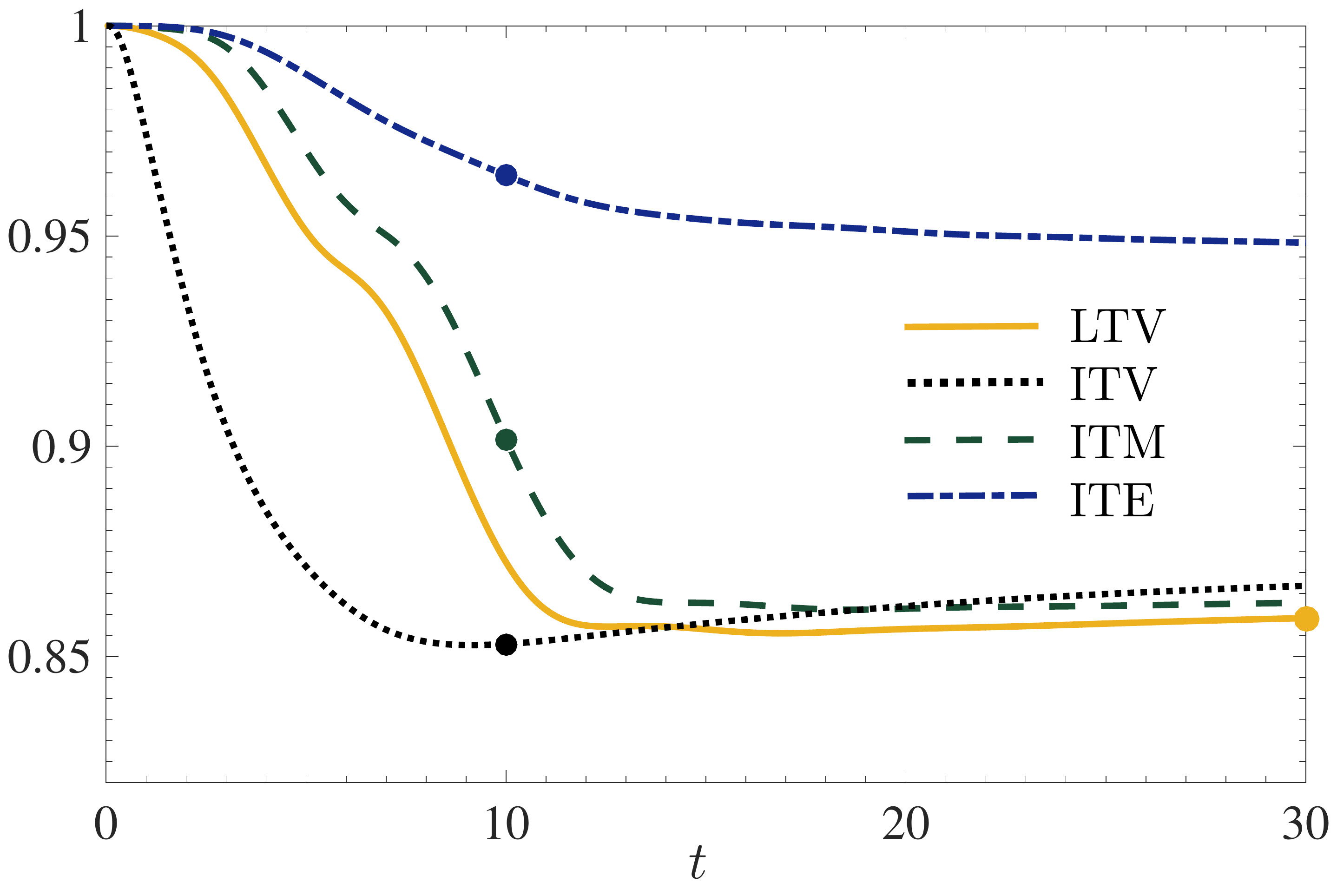}
\put(85,58){\colorbox{white}{\parbox{0.02\linewidth}{b)}}}
\end{overpic}
\captionsetup{width=0.9\textwidth}
\caption{\small{\tcr{
\tcr{For flows with: a) $Ri_b=0.1$; b) $Ri_b=0.5$};
time-evolution of the scaled variance $V$ (as defined in (\ref{defV}))
for different optimisation problem flows  where two target times $T$
have been considered for comparison: long-time ($T=30$) variance
minimisation flow (LTV; \tcr{plotted with a} solid line),
intermediate-time ($T=10$) variance minimisation flow (ITV;
dotted line), intermediate-time mix-norm minimisation flow (ITM; dashed line) and intermediate-time energy maximisation flow (ITE; dotted/dashed line). For each case, the target time for optimisation is highlighted with a colour-filled circle.}}}
 \label{fig:VvsT}
\end{center}
\end{figure}

In figure  \ref{fig:VvsRi}, we plot  
the normalised variance  $V(30)$ achieved at
time $t=30$ for \tcr{three} different optimisation problem flows,
\tcr{all of which are calculated over what we refer to as the `intermediate' time horizon $[0,10]$.
The results for the flows that 
 maximise the time-averaged perturbation kinetic energy  are plotted with a dashed  line,
the results for the flows that
minimise the mix-norm  are plotted with a solid line,} \tcr{
while the results for the flows that minimise the variance over the
same intermediate time horizon $[0,10]$ are given for reference and plotted
with star symbols.}
For this measure, it is clear that the perturbations 
which minimise the mix-norm 
are  significantly more effective than the perturbations which
maximise the time-averaged perturbation kinetic energy at identifying flows
which  enhance the reduction in the scalar variance through advection,
whatever the degree of stratification. 
 Furthermore, although (stable) stratification always tends to inhibit
 homogenisation significantly (in that $V(t)$ increases with $Ri_b$),
the transition towards poor homogenisation  (i.e. close to purely
diffusive homogenisation, with the scaled variance $V(t)
\simeq 1$) occurs at much higher
Richardson numbers for the 
flows arising from the mix-norm minimisation problem. 

In the flow arising from the time-averaged-kinetic-energy-maximisation
problem, 
$V(t)$ essentially reaches a plateau as soon as the
filamented structures vanish (with $V(30) \sim 0.9$ at
$Ri_b=0.2$), implying  that flow-induced mixing is largely ineffective
for stronger stratifications for such perturbations. These
observations extend the results of \cite{FCS14}, who, considering a
passive scalar, 
demonstrated that perturbations which maximise the time-averaged
perturbation kinetic
energy are significantly suboptimal in minimising the scalar
variance over finite time horizons. We find that this statement holds
true for an \emph {active} scalar, and is even made stronger in 
the presence of a statically stable stratification. 

Furthermore, the results presented in figure \ref{fig:VvsRi} \tcr{and \ref{fig:VvsT}} provide further evidence
that the mix-norm is a robust, useful and computationally attractive
measure to use in mixing optimisation problems, even when there
are buoyancy forces involved.Indeed, by comparison of the curves for the variance-minimisation
flows and the mix-norm-minimisation flows, the variance for the
mix-norm minimising flows is typically slightly {\itshape smaller} at the late
time $t=30$, except for the extreme flow with $Ri_b=1$, where the flow is ineffective at mixing, and the difference is very small. This does not demonstrate a break down in the optimisation
procedure of
course,
since our calculations are minimising the measure of interest over 
intermediate times (i.e. $T=10$), while the figure plots the subsequent
behaviour of the flow at late times (i.e. $t=30$).

\tcr{Entirely consistently with
the results of \cite{FCS14}, initial optimal perturbations
which lead to minimisation of variance over relatively
short times are non-trivially different from optimal perturbations
which lead to variance-minimisation at long times, typically
because such shorter time optimal perturbations
are relatively fine-scale, and so do not lead to 
thorough mixing across the channel. On the other hand, the 
mix-norm minimisation optimal perturbations
over intermediate times are close approximations 
to the variance-minimising optimal perturbations over long times,
although their computational calculation is much shorter for two
reasons. First, the loops at the heart of the DAL method are obviously
shorter when optimising over shorter target times. Second, 
 the number of iterations around such loops are
also typically and substantially
smaller, as values of the mix-norm decrease 
more rapidly in time than the values of the variance. This second phenomenon
can be appreciated through consideration of the (nondimensional) 
forms of the two evolution equations
for these measures (\ref{dVdt}) and (\ref{eq:mixevoleq}).
For the variance, the term on the right-hand-side of (\ref{dVdt}) 
is scaled with $1/Pe$, and so in general inevitably decays
slowly for large $Pe$, while the first term on the right hand side of 
(\ref{eq:mixevoleq}) 
typically ensures that the mix-norm decays more rapidly.}

\tcr{Significantly, we find that this usefulness of mix-norm minimisation
over intermediate times as a computationally efficient and robust 
proxy for variance minimisation over longer times also occurs for 
flows with all values of the bulk Richardson number which we have
considered.
This is a major result of this study, as it demonstrates that 
use of the mix-norm remains appropriate for flows
where the problem of interest is the optimisation of the
(irreversible) mixing of a {\itshape active} scalar.
To illustrate this observation further, in 
figures \ref{fig:VvsT}a and \ref{fig:VvsT}b for flows
with $Ri_b=0.1$ and $Ri_b=0.5$ respectively, we plot the 
 time-evolutions of the scaled variance for four different
 optimisation problems. In each case, the key comparison is with the evolution
of the variance for the variance-minimisation flow over the relatively
long time interval $T=30$, labelled `LTV', and plotted with a solid line.
It is clear that  minimisation of the mix-norm with the `intermediate'
target time $T=10$ (labelled `ITM', and plotted with a dashed line)
leads to significantly more mixing at time $t=30$ than the
variance-minimisation
flow over the same intermediate target time $T=10$ (labelled `ITV',
and plotted with a dotted line),
and even yields similar variance at time $t=30$ to the LTV flow.
}

\tcr{Note at the intermediate target time $T=10$, 
the variance of the ITV flow is indeed, in both cases, smaller
than the variance of the ITM flow, as expected.
It is perhaps counter-intuitive that there is a slight increase in the
scaled variance evolution for ITM, ITV and LTV flows at $Ri_b=0.5$.
It is important to appreciate that these quantities are scaled by the
(time-dependent) variance of the density distribution
experiencing pure diffusion  (i.e. with no macroscopic velocity ${\bf
  u}={\bf U}={\bf 0}$). 
At early times, the various optimal solutions develop very strong
gradients, rapidly suppressed by diffusion, 
and so the variance drops very quickly. At later times therefore, the rate
of decrease of variance of a close-to-homogeneous distribution can be extremely slow, indeed slower
than the rate associated with the purely diffusing solution, and so
the scaled variance can increase (slightly).}

The apparent particular robustness of mix-norm-based optimisation \tcr{is due} to the
mix-norm mathematical definition, which (as mentioned in the
introduction) favours filamentation of the scalar field. Indeed,
variance-based optimisation leads to the suppression of all
heterogeneities by the time horizon, whereas mix-norm-based
optimisation favours creation of small scales and sharp, filamented
structures in the scalar field. These structures are then left for
diffusion to smooth them out even after the target time, leading to
further decrease of the scalar variance. \tcr{Optimisation of the variance
over too short a target time does not ensure thorough mixing
throughout the entire width of the channel. Therefore, even for
stratified flows, we believe it is appropriate to analyse in further
detail the properties of the results from the mix-norm minimisation ITM flows.}


\section{Energy reservoir exchange}


\subsection{Available and background potential energy}

A major distinguishing characteristic of mixing of active scalars in
stratified fluids is that there is an energetic cost to the mixing in
that the potential energy of the system is ultimately increased
irreversibly when mixing occurs in an initially statically stable
fluid.
This fundamental characteristic has profound effects on the dynamics of stratified
mixing, and continues to be the topic of much ongoing research and
controversy. 
Taking the appropriate volume-averaging scalar product of the Navier-Stokes equations (\ref{NS})
with the velocity field $\bf u$ yields, after integration by parts and
application of various boundary conditions, we obtain
the evolution equation for the perturbation kinetic energy density $K=\tfrac{1}{2} \| {\bf u} \| ^2_2$:
\be
\label{dtu2}
\fb{dK}{dt} \ = \ - \fb{Ri_b}{V_\Omega}  \int_\Omega \rho v \, d\Omega \
- \   \fb{1}{V_\Omega} \int_\Omega u v \fb{\pd U}{\pd y} \, d\Omega \
- \ \fb{1}{Re V_\Omega}  \int_\Omega \nabla {\bf u}^T : \nabla{\bf u} \, d\Omega.
\ee
The various contributions on the right-hand side are readily
identified \tcr{with various key physical processes,} and we rewrite (\ref{dtu2}) in the more compact form
\be
\label{dtK}
\fb{dK}{dt} \ = \ {\cal B} \, + \, {\cal P} \, + \, {\cal D},
\ee
where ${\cal B}$ denotes the buoyancy flux exchange between the kinetic
energy and the potential energy reservoirs, ${\cal P}$ \tcr{denotes} the Reynolds
stress-mediated energy transfer rate from the pressure-driven base
flow $\bf U$ to the perturbation field $\bf u$, and ${\cal D}$ \tcr{denotes} the
\tcr{negative definite} viscous dissipation rate of conversion of perturbation kinetic energy
into internal energy.
Once again, it is important to remember that there is no assumption
that the perturbation velocity here is small compared to the base
flow, or indeed that it has zero streamwise mean instantaneously.

Similarly, the evolution equation for the potential energy density $P=
\fb{Ri_b}{V_\Omega}\int_\Omega (\rho y)  \, d\Omega$ (associated
with the dimensionless density deviation $\rho$)
can be derived from (\ref{drhodt}):
\be
\label{dPdt}
\fb{dP}{dt} \ =\ \fb{Ri_b}{V_\Omega} \int_\Omega \rho v \, d\Omega \ +
\ \fb{Ri_b}{Pe V_\Omega} \int_\Omega \left (\nabla^2 \rho \right
  ) y \, d\Omega \ =
\ - \, {\cal B} \,+ \, {\cal D}_\rho,
\ee
where ${\cal D}_\rho$ denotes the conversion rate of internal energy into
potential energy by diffusion of a statically stable density
distribution.
 Exchanges between the kinetic energy and potential energy reservoirs
 are possible through the buoyancy flux contribution in
 (\ref{dtu2})-(\ref{dPdt}). These exchanges \tcr{typically have both 
reversible and irreversible components.}
 \cite{L55} first
 introduced the concept of \textit{available potential energy} in
 order to refer to the fraction of potential energy which can be
 converted back into kinetic energy by buoyancy flux. In the absence
 of any buoyant motion, i.e. if the stratification is statically
 stable everywhere and the density gradient is parallel to the gravity
 field, this available potential energy is therefore zero. The
 remaining fraction \tcr{is conventionally  labelled the} \textit{background
   potential energy} (see for example \cite{W95}, and \cite{PC03} for
 a review) and is notionally defined as the minimum potential energy
 obtainable by adiabatic spatial redistribution of the density field
 (which forms the so-called background state).

 This partitioning of
 total potential energy into available and background potential energy
 is a convenient way of distinguishing between two different
 situations. In the first, completely reversible situation, the
 displaced fluid particles fall back to their neutrally buoyant
 position and the potential energy stored in their initial
 displacement is entirely converted back into kinetic energy. This
 may be said to correspond to  `stirring' of the scalar field by the velocity field. In the second situation,
fluid motion modifies the density field at sufficiently small scales
for diffusive fluxes to be enhanced, such that the background state is
modified, increasing the background potential energy and irreversible
`mixing' occurs. In order to account for the enhancement of diffusive
fluxes by stirring (compared to the purely conductive case), we can
thus define,
following \cite{CP00} and \cite{PC03} the irreversible (advective)
mixing rate ${\cal M}$, which acts as a sink in the evolution equation for
the available potential energy $P_A$ and as a source for the
background potential energy $P_B$ (remembering that the total
potential energy density $P=P_A+P_B$ by construction):
\bme
\label{dtPB}
\be
\fb{dP_A}{dt}  \ = \ - \, {\cal B} \, - \, {\cal M}, \qquad \qquad
\fb{dP_B}{dt}  \ = \ {\cal M} \, + \, {\cal D}_\rho.
\ee
\eme
\tcr{As in \cite{CP00}, the instantaneous background potential energy
  $P_B$ is determined at any time by spatially redistributing the
  density field using a sorting algorithm. }
\tcr{It is possible to calculate ${\cal D}_\rho$ directly from
  (\ref{dPdt}), and then  ${\cal M}$ can be determined from (\ref{dtPB}).} To understand  the actual irreversible mixing processes which 
occur in the flows we are considering,  it is
therefore illuminating to analyse the evolution of the energy stored
in the different reservoirs as the flows determined by the two
different optimisation problems evolve. Note
that, differently from \cite{W95}, we distinguish between the
irreversible mixing
inherently due to motion, occuring at a rate ${\cal M}$, and the increase
in $P_B$ that would occur in the absence of macroscopic motion, occuring at a rate
${\cal D}_\rho$.
\tcr{It is also important to remember that since} we are considering an incompressible Boussinesq fluid with a linear equation 
of state, the true thermodynamics
of stratified mixing and the exchanges between the different 
reservoirs  have been significantly simplified. These important
aspects have been reviewed in detail in 
\cite{T2013}.


\subsection{Time evolution of energy reservoirs}

\begin{figure}
\includegraphics[width=1.1\textwidth,center,trim=0 40 0 0,clip=true]{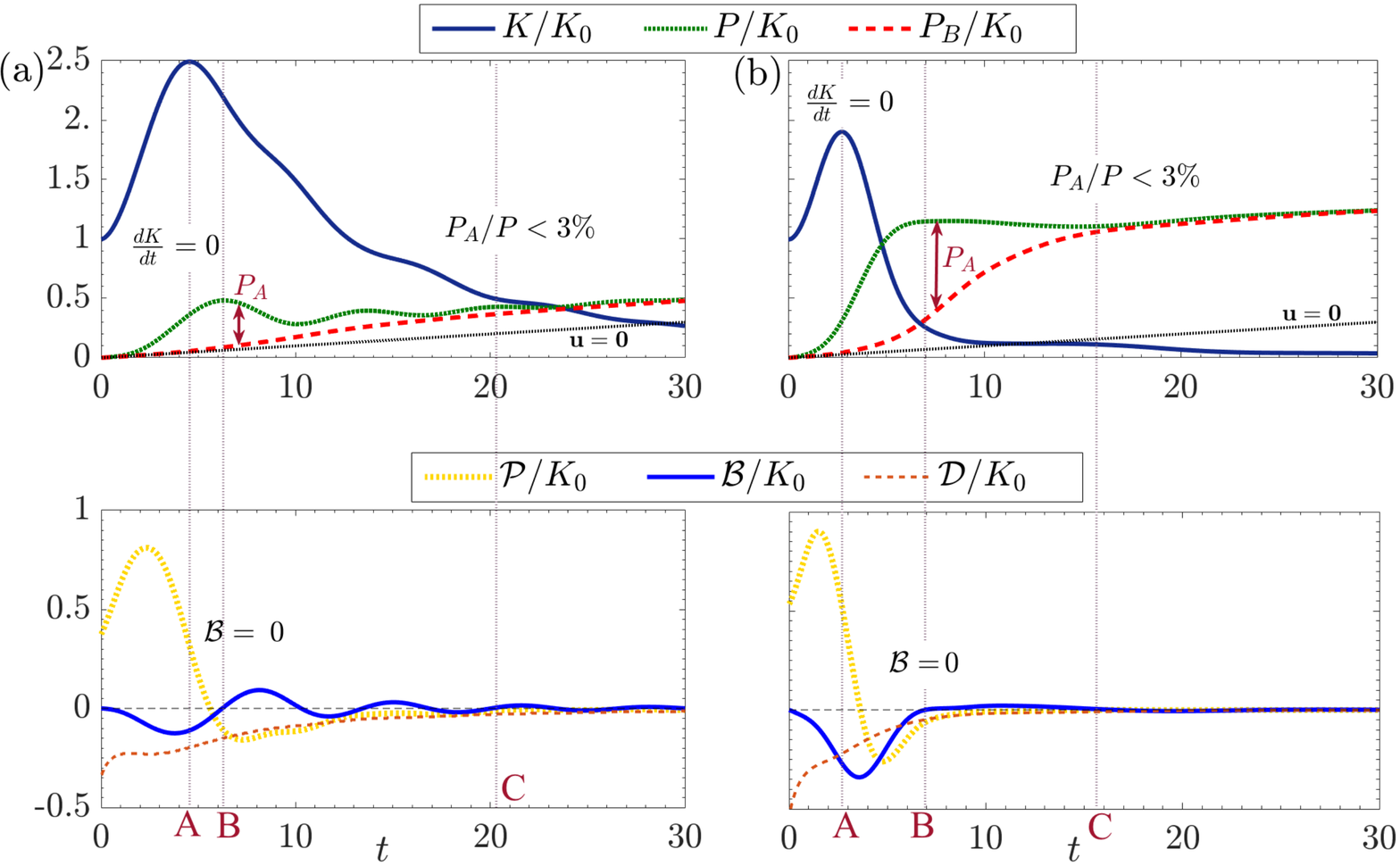}
\captionsetup{width=0.9\textwidth}
\caption{\small{Variation with time of:  (upper panels)
scaled kinetic energy density $K/K_0$ (plotted with a solid line),
scaled total potential energy $P/K_0$ (dotted line), scaled background
potential energy density $P_B/K_0$ (dashed line), and scaled potential
energy of the purely diffusing solution $\rho_d(y,t)$  (thinner
dotted line);
(lower panels) the scaled buoyancy flux ${\cal B}/K_0$ (plotted
with a solid line), viscous dissipation rate ${\cal D}/K_0$ and  energy
transfer rate ${\cal P}/K_0$ (dotted line), as defined in
(\ref{dtu2})-(\ref{dtK}).
The data comes from simulations of 
 flows with $Ri_b=0.1$ associated 
with initial perturbations which: (a) maximise the 
time-averaged perturbation kinetic energy; 
(b) minimise the mix-norm. In the upper panels, the available
potential energy $P_A$ corresponds to the difference between the
(thicker) dotted lines and the dashed lines. The following dimensionless times, which correspond to the specific events indicated on the plots and discussed in the text, are given for reference: `A'$=4.5$, `B'$=6.2$, `C'$=20.3$ (time-averaged-kinetic-energy-maximising case) ; `A'$=2.7$, `B'$=6.9$, `C'$=15.7$ (mix-norm-minimising case).}}
\label{fig:enEM}
\end{figure}

In what follows, the two types of initial 
perturbations, determined by considering the \tcr{`intermediate'} time horizon $T=10$ 
for the time-averaged-kinetic-energy-maximisation problem
and the mix-norm minimisation problem,
are added to the pressure-driven base flow $U(y)$, defined in
(\ref{eq:baseu}),
and initial 
density deviation $\rho_i(x)$, defined in (\ref{eq:rhotzero}).
These initial conditions are then used to integrate the flow forwards
to \tcr{the `long'} 
 time $t=30$. We calculate the  kinetic energy density, the total
 potential energy density and the background potential energy density
 at each time step. We plot the time evolution of these energies in
 figure \ref{fig:enEM} for the 
two different flows with $Ri_b=0.1$, along with the contributions of
the different terms \tcr{as defined} in (\ref{dtK}). The evolution of the potential
energy (equivalent in this specific case to the background potential
energy) of the purely diffusive solution 
$\rho_d(\tcb{y},t)$
(with $\bf u = \bf U = \bf 0$) used to scale the normalised variance $V(t)$, 
has been superimposed for reference (\tcr{plotted with thinner dotted
  lines in the upper panels    of figure \ref{fig:enEM}}).

\tcr{As can observed from the time-evolution of buoyancy flux in
  figure \ref{fig:enEM}, all the dynamically important buoyancy-driven motions have 
clearly 
finished significantly before the chosen final time $t=30$ for the flow with $Ri_b=0.1$. 
It is important to 
remember 
that the key (dimensional) time scale of buoyancy-driven motions is proportional 
to $\sqrt{\bar{\rho}^*/(g^* \rho^*_0)}$, and so, in the chosen
nondimensionalisation scheme, the characteristic buoyancy time scale 
is inversely proportional to the square root of the bulk Richardson 
number $Ri_b$. Therefore, we believe that using $t=30$ as the
final 
time for our various analyses is appropriate to capture important
buoyancy-driven processes for the various stratified flows considered
here,
particularly for the flows with $Ri_b \geq 0.1$ where
buoyancy becomes increasingly more significant.}

The time evolution 
develops through four phases, divided on the figures by
vertical dot-dashed lines. There is a clear relationship
to the three stages of `transport', `dispersion' and `relaxation'
discussed in the context of  passive scalar mixing by \cite{FCS14}, 
although the possibility of  exchange 
between the kinetic energy and potential energy reservoirs makes the
behaviours
of these stratified flows inevitably more complicated. 
During the first phase (up to the instant labelled `A' on figure
\ref{fig:enEM}), 
the kinetic energy $K$ experiences transient growth as energy is
transferred from the base flow to the perturbation at \tcr{the
  `production'} rate ${\cal P}$. 
This stage has a longer duration in the flow associated
with the energy-maximisation problem (left \tcr{column})
 than in the flow associated with  mix-norm minimisation (right
 \tcr{column}), and results in a significantly higher kinetic energy at its 
 peak. This is unsurprising, since perturbation energy is 
of course being maximised in one flow and not in the other.
This phase  corresponds to an initial part of the  `transport' phase of 
mixing, where much of this growth of
perturbation kinetic energy can be associated
with extraction of kinetic energy from the base flow via the Orr
mechanism.

In the second phase, (between the instants marked `A' and `B')
the kinetic energy decreases and ${\cal B} < 0$. Therefore,  
the energy transfer rate from the base flow is no longer sufficient to
counteract the combined effects of the viscous dissipation rate ${\cal D}$
converting 
kinetic energy into internal energy and the 
buoyancy flux ${\cal B}$ converting kinetic energy into potential energy.
During this phase, since ${\cal B} < 0$, the potential energy increases.
The energetic perturbation 
 is lifting up
dense fluid and pushing down light fluid to create the
vertical 
`stripe', rapidly distorted to `chevron' structure conducive to subsequent 
mixing, and characteristic of a later part of the initial `transport'
stage.

Both $P_A$ and $P_B$ increase for both flows, 
but there is substantial qualitative difference between the two
flows. 
In the flow associated with (kinetic) energy-maximisation, 
$P_B$ increases only slightly faster than the purely 
diffusive solution (plotted with the thinner dotted line), as $M$ is
very small in magnitude, 
and the perturbation kinetic energy does not actually decrease
substantially.
Even more (ultimately) significantly,
the total potential energy $P$ (plotted with the thicker dotted line) grows to a much smaller maximum value than in the flow 
associated with mix-norm minimisation. In the mix-norm minimisation
flow, 
$P_B$ (plotted with a dashed line) also grows much more strongly during this stage, (associated
with a much larger value of $M$, the advective irreversible mixing
rate). For this flow, the kinetic energy has decreased much more
significantly, and the flow arrangement appears to be consistent with
the perturbation kinetic energy `efficiently' being exchanged into
potential energy.

The second phase finishes (at `B') when the total potential energy
density $P$ reaches a (local) maximum, and equivalently when the
buoyancy flux ${\cal B}=0$. 
The third (and longest) phase (between the times marked `B' and `C') 
is the phase during which the majority of the  irreversible mixing occurs, with 
$P_B$ increasing and $P_A$ decreasing, until
the total potential energy and the background potential 
energy are within $3\%$ of each other.  This phase
naturally corresponds to the second `dispersion'
phase of (significant) mixing, while the fourth phase (when $P_B \lesssim P$) for 
times after `C' clearly 
corresponds to the final diffusion-dominated `relaxation' stage
discussed
by \cite{FCS14}. 

As can be seen  in the left column of  figure \ref{fig:enEM}, in the 
flow associated with the time-averaged-kinetic-energy-maximisation problem,
 the background potential energy increases monotonically with 
almost equal contributions from the (notional) diffusive conversion of internal 
energy and the decrease of available potential energy. Furthermore,
for this flow, the total potential energy varies non-monotonically,
with the buoyancy flux mediating a quasi-periodic oscillatory 
exchange between the kinetic energy and the potential energy
reservoirs.

The behaviour of the flow associated with the mix-norm minimisation
problem during this significant mixing phase (or equivalently the
`dispersion' stage) is qualitatively
different, with a much larger increase in the background potential
energy. Furthermore, there 
is a very marked damping of subsequent buoyancy flux oscillations.
Essentially, virtually all the energy stored as 
available potential 
energy by the development of the perturbation during the first two
phases (i.e. during the `transport' stage either due to kinetic energy increase
or potential energy increase)
of the flow evolution is 
converted into background potential energy, and thus is associated
with irreversible mixing during this phase. 

Therefore, there are two
key characteristics of this flow which appear to be conducive to mixing.
First, a relatively large amount of energy is converted into potential
energy. Second, virtually all of this energy is `captured' as
background potential energy through irreversible mixing in a single 
pass, without oscillatory, and hence at least partially reversible,
exchange with the kinetic energy reservoir. Two natural questions then
arise. First, can these two characteristics be identified with particular
physical 
processes identifiable in the flow evolution? Second, 
do these characteristics lead to significant differences in the \emph
  {efficiency}
of the mixing in an energetic sense \tcr{(as defined in \ref{ame})}, analogous to that
defined in the context of transient shear instability-driven stratified
mixing considered by \citet{W95} and \citet{CP00}?


\subsection{Flow evolution and physical interpretation}

\begin{figure}
\includegraphics[width=\textwidth]{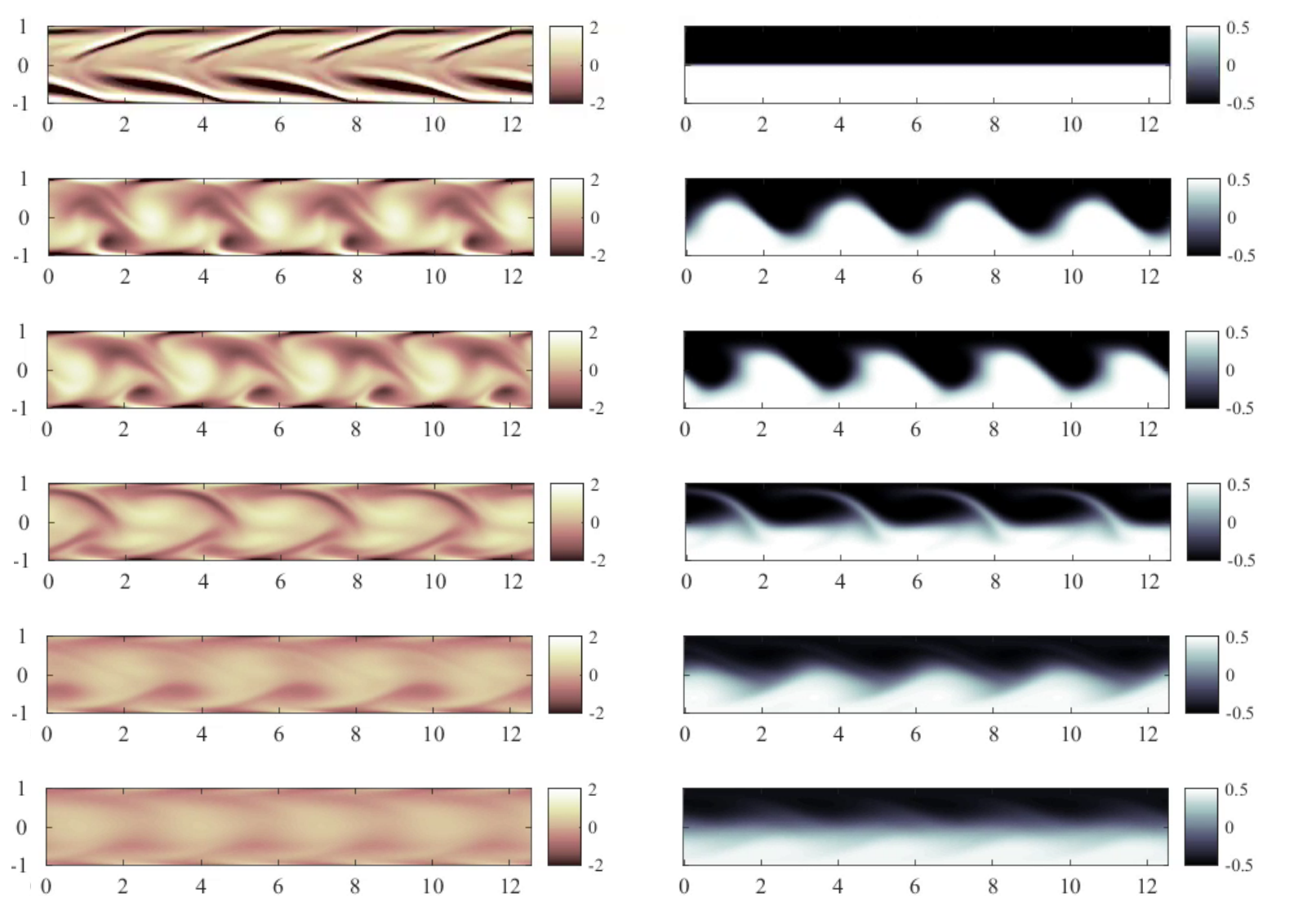}
\captionsetup{width=0.9\textwidth}
\caption{\small{Snapshots of the vorticity field \textit{(left
      column)} and density field \textit{(right column)} for the flow
    induced by 
the optimal perturbation for time-averaged perturbation kinetic energy maximisation
with $Ri_b=0.1$ at times as marked in the left column of figure \ref{fig:enEM}:  from top to bottom, $T=0$;  $T=A=4.5$;  $T=B=6.2$;  $T=10$; $T=C=20.3$; $T=30$.}}
\label{fig:snapRi01E}
\end{figure}

\begin{figure}
\includegraphics[width=\textwidth]{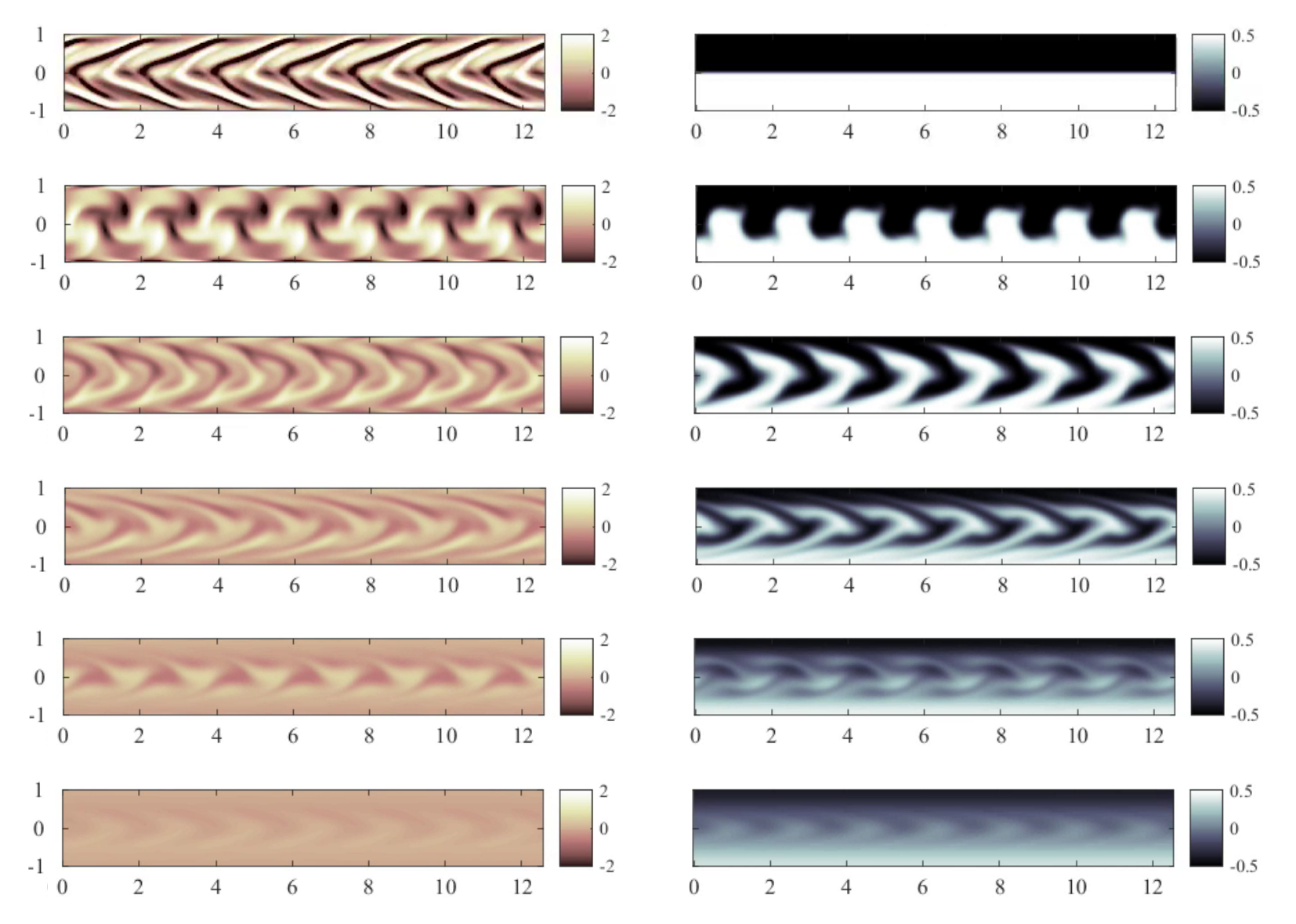}
\captionsetup{width=0.9\textwidth}
\caption{\small{Snapshots of the vorticity field \textit{(left
      column)} and density field \textit{(right column)} for the flow
    induced by 
the optimal perturbation for  mix-norm minimisation
with $Ri_b=0.1$ at times as marked in the right column of figure \ref{fig:enEM}:  from top to bottom, $T=0$;  $T=A=2.7$;  $T=B=6.9$; $T=10$; $T=C=15.7$; $T=30$.}}\label{fig:snapRi01M}
\end{figure}

The four different phases  in the evolution of the energy reservoirs
can be identified with a succession of flow 
structures plotted in figures \ref{fig:snapRi01E} and
\ref{fig:snapRi01M}, where the  perturbation vorticity field and
density deviation field are shown. In both flows, the first peak in
total potential energy (marked as `B') is achieved when the initially
prescribed vorticity field has distorted the interface maximally, and
brought light fluid to a minimal height and 
dense fluid to a maximum height, completing the  initial transport
stage. As expected, 
the vorticity field also shows  a clear signature of the Orr mechanism 
between $t=0$ and $t=$`A' when the perturbation kinetic energy is
maximum. 
At $t=0$, 
the vortices are initially elongated and tilted against the base flow
shear, then shrink to a more compact shape (causing the kinetic energy
to temporarily increase) at $t=$`A', 
 before the base flow stretches the vortices further and inclines them
 in the opposite direction, as is apparent by $t=10$ (corresponding to
 the optimisation time horizon) in both flows.

Furthermore, in both flows, 
 most of the conversion from available potential energy to background
 potential energy occurs between the times marked `B' and `C', 
subsequent to which the effects of advection-driven irreversible
mixing  becomes negligible compared to those of purely molecular
diffusion. 
However, the striking difference between the two flows  is that
in the 
flow initialised by the optimal perturbation for mix-norm
minimisation, (shown in figure \ref{fig:snapRi01M}) 
 the perturbation kinetic energy is markedly smaller.  The
 density field 
consists of elongated protrusions, tilted by the base flow velocity field
to project  dense and light fluid filaments  into the lower  and upper half-planes
respectively, aligned with the base velocity. The
irreversible mixing occuring afterwards therefore mostly relies on
stirring by the base flow rather than the perturbation, taking
advantage of Taylor dispersion.
It is clear from the snapshots in figure \ref{fig:snapRi01M} that from
time `B', the structure of the density  does not experience any
significant structural change, but is rather  advected  by the base
flow and smoothed out by diffusion.  

Conversely, for the flow associated
with the initial
perturbation
which maximises the time-averaged perturbation kinetic energy, 
the evolution of the density field during the third phase of flow
evolution (i.e. the `dispersion' stage)  is 
strongly affected by 
stirring (by its nature at least partially reversible) 
by the still non-trivially
energetic
perturbation velocity field. This perturbation 
is central to the 
creation of the relatively small filamented structures visible in
figure \ref{fig:snapRi01E}. 
Remarkably, the inflection point in the time evolution of the
background potential energy in figure \ref{fig:enEM}a lies between
times $t=10$ and $t=15$, indicating that the irreversible mixing rate
$M$ reaches its maximum in this interval, interestingly slightly after
the optimisation time horizon $T=10$. The snapshots shown in
figure \ref{fig:snapRi01E} 
demonstrate that this maximum in the mixing corresponds to the roll-up
of these thin, and sparsely distributed protruding filaments becoming subject to the restoring buoyancy force.


\subsection{Assessing mixing efficiency}
\label{ame}
\begin{figure}
\begin{center}
\includegraphics[height=0.31\textwidth]{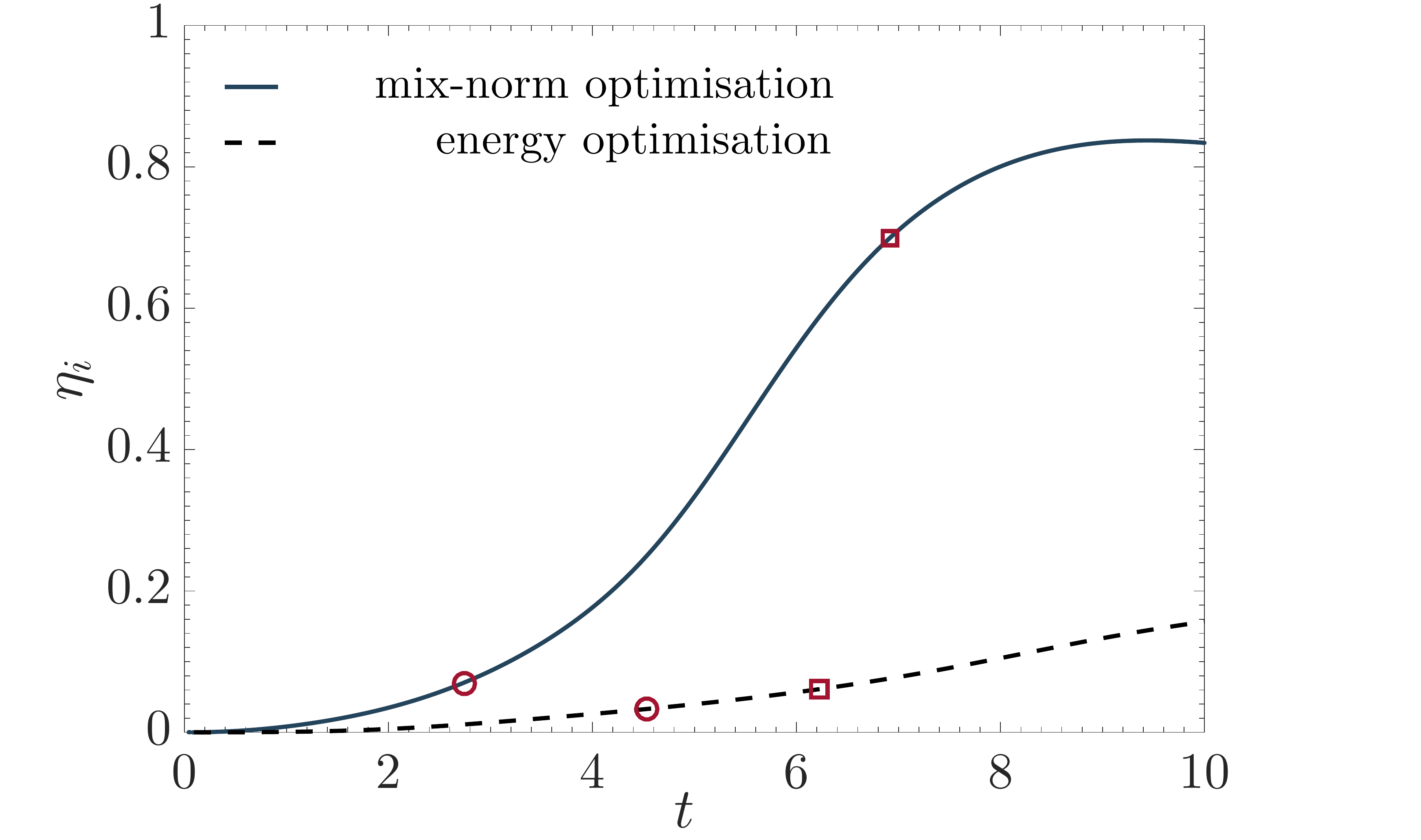}
\hspace{-7ex}
\includegraphics[height=0.31\textwidth]{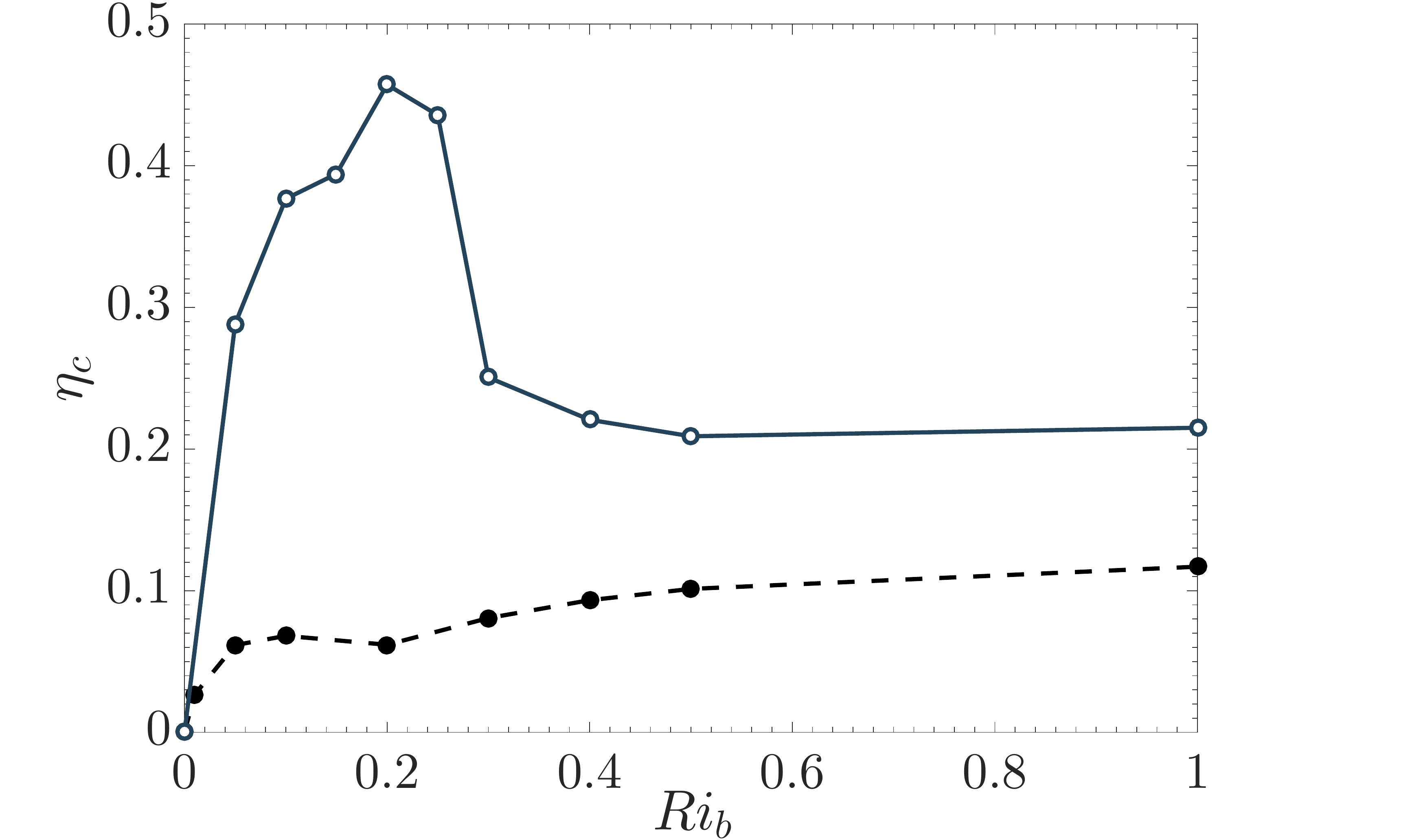}
\end{center}
\captionsetup{width=0.9\textwidth}
\caption{\small{\textit{Left:} Time-evolution of the instantaneous
    mixing efficiency $\eta_i$ (as defined in (\ref{eq:etai})) over the optimisation  time horizon
    $[0,10]$, for  flows with $Ri_b=0.1$ initialised with optimal
    perturbations for mix-norm minimisation (plotted with a solid
    line)
and for time-averaged-kinetic-energy maximisation  (dashed line). In both cases times `A' (circles) and `B' (squares) are indicated for reference.
\textit{Right:} Variation with $Ri_b$ of the cumulative mixing
efficiency $\eta_c$ (as defined in (\ref{eq:etac})) integrated up to
time $T^*=30$, 
for flows initialised with 
perturbations
for 
mix-norm minimisation (solid line and empty circles) and
time-averaged-kinetic-energy maximisation (dashed line and full circles).}}
\label{mixeff}
\end{figure}

Following \citet{CP00} and \citet{PC03}, we can now use the inferred
irreversible mixing rate $M$ introduced in (\ref{dtPB}) to define an
appropriate instantaneous mixing efficiency 
\be
\eta_i=\frac{{\cal M}(t)}{{\cal M}(t)-{\cal D}(t)}, \label{eq:etai}
\ee
the ratio of the irreversible mixing rate inherently
due to fluid motion to the sum of this rate and the viscous
dissipation of the perturbation kinetic energy. 
This instantaneous mixing efficiency, quantifying as it does
the proportion of kinetic energy lost by the perturbation 
that leads to irreversible mixing,  has the further attraction of
being easily computed at each time step.
We plot its time evolution  in figure \ref{mixeff}a as a function of
time over the optimisation time interval $[0,10]$, for the same flows
with $Ri_b=0.1$ 
associated with mix-norm minimisation  (plotted with a solid line) and 
the flow associated with maximisation 
of the time-averaged
perturbation 
kinetic energy (dashed line). 

For the 
flow associated with mix-norm minimisation, 
$\eta_i$ 
 steeply increases between the times `A' and `B' as the total
 potential \tcr{energy} reaches a plateau at the end of the transport phase,
associated with (clearly) efficient 
exchange of kinetic energy 
to the potential energy reservoir, coincident also with  the available
potential energy reservoir emptying  and the background potential
energy reservoir increasing. 
For the flow associated
with time-averaged-kinetic-energy maximisation,
 the increase in $\eta_i$ is significantly more modest, reaching
an appreciably smaller
peak value, 
illustrating in a different way
that 
this optimisation 
problem is 
suboptimal 
 in causing mixing in such a stratified flow.

This picture appears to be largely consistent for 
all the flows we have considered with different $Ri_b$. To demonstrate
this we calculate an appropriate \emph{cumulative}
mixing efficiency $\eta_c$, defined as
\be
\eta_c=\fb{\int_0^{T^*} {\cal M}(t) \, dt}{\int_0^{T^*} \left [{\cal
      M}(t) - {\cal D}(t) \right ] \,
  dt}=\fb{P_B(T^*)-P_d(T^*)}{P_B(T^*)-P_d(T^*)-\int_0^{T^*}
  {\cal D}(t) \, dt},
\label{eq:etac}
\ee
%
%
where $P_d(t)$ is the potential energy density associated with the
purely diffusive density field $\rho_d$. 
We plot $\eta_c$ as a function of $Ri_b$ in 
 figure \ref{mixeff}b 
for the two classes of flows \tcr{on which we are focussing}: those 
associated with mix-norm minimisation (plotted with a solid line);
and those associated with time-averaged perturbation kinetic energy maximisation (dashed
line). These optimisation 
problems \tcr{have all been calculated with respect to the `intermediate'} time horizon $T=10$, 
but we set the terminal time in the calculation 
of $\eta_c$ to be \tcr{the `long' time} $T^*=30$. 

It should be emphasized that, since ${\cal M}$ is proportional to $Ri_b$, 
mixing efficiency defined in this way inevitably tends to zero
in the limit of zero stratification.
This measure further highlights the suboptimality of the
energy-maximising flows 
for all the considered
 stratifications. 
For both optimisation problems, the cumulative mixing 
efficiency plateaus
at higher $Ri_b$, with 
$\eta_c$ for the mix-norm-minimising flows
approaching
an asymptotic value nearly twice the equivalent 
asymptotic value 
for the energy-maximising flows. 
These asymptotic values correspond 
to flows where the initial perturbation
is 
no longer sufficiently energetic
for the filamentary structures to develop adequately to form the
`stripes' or `chevrons' necessary to exploit Taylor dispersion 
to enhance diffusive mixing. Interestingly, and perhaps fortuitously,
this
asymptotic value $\eta_c \simeq 0.2$ is quite close to, \tcr{but
  slightly larger than} the classic
bounding value of \cite{Osborn1980} of the turbulent 
flux coefficient $\Gamma \simeq  \eta_c/(1-\eta_c) \leq 0.2$,
\tcr{since for this flow $\Gamma_c \simeq 0.25$. However, it
is very important to remember that this flow is restricted
to evolve in two dimensions, 
and so the particular quantitative mixing dynamics must be treated with caution.}

 On the other hand, the noticeable  kink in the solid curve for the
 mix-norm-minimising flows 
around $Ri_b=0.2$ suggests  a particularly favourable flow
configuration, consistent with the evolution of the density 
fields already discussed in \ref{sec:opt_strat}. 
 For smaller $Ri_b \lesssim 0.25$, the base flow shear prevents the
 `chevrons'  from collapsing back  to their neutrally buoyant
 positions, thus allowing  the filaments to remain in the near-wall
 regions of higher shear  where Taylor dispersion ensures efficient
 mixing. Furthermore, for  $Ri_b \simeq 0.2-0.25$, the combined
 effects of this Taylor
 dispersion and (significantly) convective overturning result in a peak in mixing
 efficiency, followed by a relatively rapid drop at slightly higher
 $Ri_b$ as the distorted interface is no longer
advected close to the boundaries, and so Taylor dispersion 
no longer occurs. 

This maximum value of the mixing
efficiency $\eta_c \simeq 0.4-0.5$ with intermediate Richardson number
is 
\tcr{similarly}
somewhat reminiscent of the behaviour of 
transient stratified shear flows prone to Kelvin-Helmholtz
instabilities (see for example \cite{Mashayek2013}), although this 
apparent agreement may of course be coincidental since
the \tcr{inherently two-dimensional} flows considered here are 
neither linearly unstable nor
turbulent at any stage.
Indeed, it is important to appreciate that the non-monotonic
dependence
of $\eta_c$ on $Ri_b$ is due to the existence of two qualitatively 
different flow dynamics at smaller and higher $Ri_b$, and 
so should not be interpreted as implying the development 
of well-mixed layers separated 
by relatively `sharp' interfaces due to the so-called `Phillips
mechanism'
\citep{Phillips1972}.


\section{Conclusions}

Using the direct-adjoint-looping (DAL) method, involving numerical integration of the fully nonlinear direct-adjoint Boussinesq
equations, we have for the first time identified initial perturbations
which lead to optimal mixing in a density-stratified  flow at
$Re=Pe=500$. For ease of comparison with \cite{FCS14}, 
the same (effectively) proof-of-concept 
 optimisation problem has been considered. 
Perturbations 
of fixed kinetic energy
in plane Poiseuille flow driven
by a constant base pressure gradient 
have been identified which 
(over the chosen intermediate time horizon $T=10$)
 either maximise the
time-averaged perturbation kinetic energy, or minimise 
 the  `mix-norm' quantifying the homogenisation of the flow.

 Consistently with the passive scalar flow results of \cite{FCS14}, 
the flows associated with both 
optimisation problems 
 exploit first an `Orr mechanism' for transferring kinetic energy from
 the base Poiseuille flow to the perturbation (leading to `transport'
 of the scalar into a `stripes' or `chevrons' structure) 
followed by  Taylor dispersion 
which leads to
enhanced diffusive fluxes before both flows eventually `relax' through
simple diffusion. Nevertheless, for all stratifications, the flows
associated with mix-norm minimisation are substantially less energetic
than 
the flows associated with energy maximisation, yet substantially
more effective at reducing the (dynamic) scalar variance. This
difference is
particularly marked for moderate values of the bulk Richardson number
($Ri_b \sim 0.2$) 
as stratification especially inhibits the development of the
large-scale, wavy structures characteristic
of flows which maximise
the time-averaged perturbation kinetic energy.

Comparison of the  time evolution of the 
kinetic energy and the various potential
energies
associated
with each of the optimisation
problems 
shows that although the transient growth in the perturbation kinetic
energy is significantly higher 
for the time-averaged-kinetic-energy-maximisation 
flows, the potential energy exhibits much larger ultimate  growth with
the mix-norm-minimisation 
flows.  This  phenomenon indicates that the development of the particular perturbations
in such flows is
much more effective
at inducing 
irreversible mixing through  modification of the initial density
distribution at small scales. 
The mix-norm-minimising optimal perturbations  prove 
reliable in identifying a particular type of stirring mechanism with
two key characteristics. First, the stirring rapidly stores
a large amount of available potential energy  once the
perturbation kinetic
energy has reached its maximum value. Second, the stirring then converts this
available potential energy, essentially in one shot, into background
potential energy, without the characteristic oscillatory, and at least
partially reversible, exchange between the kinetic energy and
potential
energy reservoirs typical of flows
which maximise time-averaged perturbation kinetic energy.

Our stratified, yet still highly idealised, results constitute further
evidence that variational optimisation problems focussed on mix-norm-minimisation
provide robust and computationally efficient indicators to identify
dynamical processes that ensure scalar mixing, even in flows with
buoyancy forces. 
\tcr{Specifically, for all the values of $Ri_b$ we consider, we 
  demonstrate 
that mix-norm minimisation over intermediate times remains 
a robust and computationally efficient proxy for 
variance-minimisation  (i.e. thorough mixing) over 
long times. This observation suggests two interesting open questions 
for future consideration. First,  how short can such an 
`intermediate' time be taken for the mix-norm minimising optimal 
perturbations still to be adequately close to the (desired) long-time 
variance-minimising perturbations? 
 A second, potentially more 
general question, is whether  
there is an `optimal' mix-norm, in the sense of a specific 
(negative) 
index of a Sobolev norm for which the associated 
minimisation problem is the most robust and computationally efficient 
proxy for identification of  long-time  optimal variance-minimising 
perturbations, 
which are typically the perturbations of most relevance.}

As
the nonlinear direct-adjoint-looping (DAL) method which we have used
here is \tcr{somewhat}
computationally demanding,  
we have here restricted our attention to a two-dimensional flow in a
simple streamwise periodic geometry, well below the linear stability
threshold, with $Pr=1$. These assumptions obviously strongly restrict 
quantitative application, but in principle the same techniques
can be used to consider three-dimensional flows at higher $Re$ and
$Pr$, as in \cite{Vermach2018}.

Indeed, this method has the potential to address some
fundamental outstanding questions in (turbulent) stratified
mixing, not least of which is the question of whether the mixing associated
with classic, and widely studied  flow instabilities (such as the Kelvin-Helmholtz
instability) is
actually `optimal'. Although there is some evidence (see for example
\cite{Mashayek2017}) that the initial large-scale overturning associated with
this
instability is particularly conducive to efficient mixing, the
methodology presented here can be straightforwardly formulated
to investigate whether other perturbations might be even more
effective at mixing. 

In a more industrial context, as also
demonstrated by \citet{FCS14}, 
the underlying optimisation problem can be formulated
to identify not an optimal initial perturbation, but rather an optimal
wall-forcing strategy to encourage mixing. What is the best forcing
strategy with a constrained power for a stratified fluid in a
constrained geometry is obviously an interesting question. Since
our results suggest that maximising perturbation 
energy may well be sub-optimal, it is at least possible that 
the optimal forcing strategy may be counter-intuitive, and 
may not rely (or indeed require) substantial injections of energy,
which would have the potential for non-trivial savings in the 
(total) energy cost required to mix fluids effectively to a required
tolerance.


\section*{Acknowledgements}
F.M. was funded by a David Crighton Fellowship from the Department of
Applied Mathematics and Theoretical Physics at the University of
Cambridge. 
The research activity of C. P. C.  is supported by the 
 EPSRC Programme Grant EP/K034529/1 entitled `Mathematical Underpinnings of Stratified Turbulence'.
Both authors wish to thank Peter J. Schmid for many fruitful
discussions and Dimitry P. G. Foures for providing a version of the
DAL method code described in \cite{FCS13}. \tcr{Furthermore, the
  constructive
comments of anonymous referees have significantly improved the clarity
and presentation of our results.} The codes and initial conditions used to generate all the data used in
the figures are available at {\tt https://doi.org/10.17863/CAM.25375}.


\bibliographystyle{jfm}

\end{document}